\documentclass[a4paper,11pt]{article}
\pdfoutput=1 
\usepackage{jheppub}

\usepackage{float}
\usepackage{gensymb}
\usepackage{subcaption}
\usepackage{soul}
\usepackage{slashed}
\usepackage{bm}
\usepackage{xcolor}
\usepackage{mathtools}
\usepackage{siunitx}
\usepackage{multirow}
\usepackage{pifont}
\usepackage{soul}
\allowdisplaybreaks[4]

\title{\boldmath Complex Scalar Singlet Model: Electroweak Phase Transition and Gravitational Waves}

\author{Dilip Kumar Ghosh, Debadrita Mukherjee, Koustav Mukherjee, and Rohan Pramanick}

\affiliation{School of Physical Sciences, Indian Association for the Cultivation of Science, Kolkata-700032, India}

\emailAdd{dilipghoshjal@gmail.com, mukherjeedebadrita531@gmail.com, koustav.physics1995@gmail.com, rohanpramanick25@gmail.com}

\abstract{The Standard Model (SM) cannot explain the observed baryon asymmetry of the Universe (BAU), thus driving the need for physics beyond the SM, which can generate the electroweak baryogenesis through a strong first-order electroweak phase transition (SFOPT). We extend the SM with a complex singlet scalar (cxSM) and examine the phase transition dynamics using a fully general renormalizable scalar potential that permits a complex vacuum expectation value for the singlet scalar and coupled dynamics among multiple scalar fields. Employing the full one-loop thermal effective potential, we conduct an extensive parameter space scan, enforcing both theoretical and experimental constraints on the scalar sector. To efficiently capture the highly non-linear structure of model parameter space, we embark on a classifier based machine learning approach to identify regions manifesting SFOPT while revealing key signatures of model parameters. From these regions, we select a few representative benchmark scenarios that show multi-stage transitions followed by the production of stochastic gravitational wave (GW) signals from bubble nucleation dynamics. The resulting spectra fall within the projected sensitivities of future space-based detectors such as LISA, BBO, DECIGO, and U-DECIGO. Thus, the cxSM offers a compelling setting for the electroweak baryogenesis, enriched by correlated gravitational wave and collider phenomenology.}

\begin{document} 
\maketitle
\flushbottom

\newcommand{\roottwo}{\sqrt{2}}
\newcommand{\roottwoinv}{\dfrac{1}{\sqrt{2}}}
\newcommand{\two}{\dfrac{1}{2}}
\newcommand{\four}{\dfrac{1}{4}}
\newcommand{\six}{\dfrac{1}{6}}
\newcommand{\eight}{\dfrac{1}{8}}
\newcommand{\hc}{\text{h.c.}}


\section{Introduction} 

The origin of the observed Baryon Asymmetry of the Universe (BAU) \cite{Planck:2018vyg} remains one of the fundamental mysteries that currently lacks a viable description within the Standard Model (SM) of particle physics, despite its tremendous success. This remains one of the key motivations for pursuing physics beyond the Standard Model (BSM), apart from the non-explanation of tiny neutrino mass and the existence of dark matter (DM). The dynamical generation of such an asymmetry requires one to meet the three Sakharov conditions, i.e, the violation of baryon and lepton number, the existence of C and CP violating processes, and the departure from thermal equilibrium \cite{Sakharov:1967dj}. The SM can account for the CP-violating processes, although their contribution turns out to be minuscule compared to observational evidence. Additionally, the electroweak symmetry breaking phase transition is a smooth crossover \cite{Kajantie:1996mn, Csikor:1998eu} with the observed Higgs boson mass, excluding the possibility of electroweak baryogenesis \cite{Dine:1992wr, Trodden:1998ym, Morrissey:2012db, vandeVis:2025efm} in the context of the SM alone.

The strong first order phase transition (SFOPT) provides a viable framework to satisfy the necessary condition for processes to go out of equilibrium. In addition, SFOPT can generate signatures of stochastic gravitational waves (GW) that can be probed in various experiments, such as the upcoming space-based interferometers, such as LISA \cite{Baker:2019nia}, BBO \cite{Crowder:2005nr, Harry:2006fi}, DECIGO, and U-DECIGO \cite{Seto:2001qf, Kawamura:2020pcg}. These offer a novel and complementary probe to investigate various BSM scenarios; for reviews, see~\cite{Mazumdar:2018dfl, Athron:2023rfq, Athron:2023xlk}.

The simplest possible scenario to achieve an SFOPT is to consider a gauge singlet scalar extension of the SM \cite{Espinosa:1993bs, Profumo:2007wc}, which has been studied extensively in the literature for various phenomenological studies, which include collider searches \cite{Braathen:2025svl}, production of GWs \cite{Ellis:2022lft}, and electroweak baryogenesis \cite{Vaskonen:2016yiu}. Apart from that, this setup also allows a possible dark matter candidate stabilized by an additional discrete or continuous global symmetry. In addition, inclusion of additional scalar doublets and other well-motivated gauge extended BSM scenarios have also been explored to achieve SFOPT \cite{Lee:2025hgb, Ghosh:2024ing, Biswas:2025rzs, Keus:2025ova, Karkout:2024ojx, Borah:2024emz, Borah:2025ubr, Guan:2025idx}.

In this work, we explore the dynamics of phase transition in a complex gauge singlet extension of the SM, considering the most general scalar potential without any additional symmetry imposition on the singlet, usually referred to as the ``cxSM''. Inclusion of non self-hermitian terms in the scalar potential with all real couplings allows a vacuum configuration with the complex vacuum expectation value (vev) of the singlet. The complex phase of singlet vev can also violate CP spontaneously in the presence of additional heavy Majorana fields \cite{Fernandez-Martinez:2022stj, Pramanick:2022put} or non-renormalizable fermionic interactions to drive electroweak baryogenesis (EWBG) \cite{Vaskonen:2016yiu}. We focus on the dynamics of the SFOPT involving three fields, i.e. the real and imaginary parts of the scalar singlet along with the SM Higgs, which serve as an essential requirement to study EWBG in cxSM, while refraining from a dedicated analysis of EWBG, which we defer to future work. The dynamics of phase transition involving multiple fields, along with the exploration of the regions of the parameter space, with a large number of input parameters, satisfying SFOPT, require extensive numerical scanning. To alleviate this issue, we implement a machine learning (ML) algorithm to find out whether a point in the parameter space can be correctly classified to yield an SFOPT. While there exists a multitude of ML techniques, in this work we restrain ourselves to the boosted decision tree approach utilizing the eXtreme Gradient Boosting (XGBoost) algorithm \cite{Chen:2016btl}. We also discuss the generated stochastic GW spectrum from the SFOPT and their detection prospects in near future detectors.

The draft is organized as follows. In section \ref{sec_model}, we describe the complex singlet extended SM, briefly mentioning the relevant theoretical as well as experimental constraints. We construct the effective potential in section \ref{sec_phase_transition} and elucidate the details of SFOPT through several benchmark points. We also investigate the possibility of implementing classifier based machine learning technique to predict the probability of exhibiting SFOPT for any point in the parameter space. The stochastic GW spectrum generated for the benchmark points is presented in section \ref{sec_gw} before finally concluding in section \ref{sec_conclusion}.


\section{Model} \label{sec_model}

The cxSM is the simplest scalar extension in which the scalar sector of the SM is augmented with a single complex gauge singlet scalar ($S$) \cite{Barger:2008jx}. To remain self-consistent, we discuss the scalar potential along with the most general vacuum configuration, highlighting all the relevant constraints on the couplings in the following subsections.


\subsection{Scalar potential}

Along with the SM Higgs doublet ($H$), the most general renormalizable potential of the cxSM at the tree level can be decomposed as
\begin{equation} \label{eq_Vtree}
    V_0 (H, S) = V_\text{SM} + V_S + V_{H S} \, ,
\end{equation}
where $V_\text{SM}$ is the SM Higgs potential, $V_S$ contains all possible cubic and quartic self interactions of $S$, while the interactions of the scalar singlet with the Higgs doublet are encapsulated in $V_{H S}$. Individual contributions are given by
\begin{equation}\label{eq_Vtree_terms}
\begin{aligned}
    V_\text{SM} &= - \two \mu_H^2 H^\dagger H + \four \lambda \left(H^\dagger H \right)^2 \, , \\
    V_S &= \left( a S + \four b_1 S^2 + \hc \right) + \two b_2 |S|^2 + \left(\six \mu_3 S^3 + \widetilde{\mu}_3 |S|^2 S + \hc \right) \\
        &+ \left( \eight d_1 S^4 + \eight d_3 |S|^2 S^2 + \hc \right) + \four d_2 |S|^4 \, , \\
    V_{H S} &= H^\dagger H  \left[ \left(\four \delta_1 S + \four \delta_3 S^2 + \hc \right) + \two \delta_2 |S|^2 \right]   \, .
\end{aligned}
\end{equation}
The linear term can be eliminated by a constant shift in the scalar field $S$; however, we keep all terms without further redefinition of cubic and quartic terms upon field shift \cite{Espinosa:2011ax}. Additionally, imposition of various discrete or continuous global symmetries on the additional scalar singlet required for different phenomenological studies of interest can drastically reduce the number of terms. Nevertheless, we work with the full potential given in Eq.~\ref{eq_Vtree} without the imposition of any \textit{ad hoc} symmetry by hand. The coefficients of the non-self hermitian terms arising in Eq.~\ref{eq_Vtree_terms} can, in general, be complex, leading to the most generic scalar potential; however, we choose all the couplings to be real for simplicity. Additionally, this choice ensures CP symmetry of the Lagrangian in the presence of additional (higher-dimensional) operators involving fermions, leveraging the possibility of spontaneous CP symmetry breaking. It is important to note that the scalar potential defined in terms of $H$ and $S$ is different from that of two real scalar singlet extensions due to the inherent $\mathbb{Z}_2$ symmetry under which the complex scalar transforms as $S \rightarrow S^\prime = S^*$ \footnote{The spontaneous breaking of this discrete symmetry leads to the formation of domain walls potentially jeopardising the homogeneous background \cite{Blasi:2022woz}. This can be avoided by introducing a tiny bias term ($\sim \mathcal{O}(10^{-6}) \text{\ GeV}^4$) in the potential which can result in a prompt decay of the domain wall without having any significant effect on the phase transition and GW parameters \cite{Pham:2024vso, Pham:2026bip}.}.

In terms of their components, the scalar fields can be represented as
\begin{equation} \label{eq_field_components}
    H =  \begin{pmatrix} G^+ \\ \dfrac{1}{\sqrt{2}} (h + i G^0) \end{pmatrix} \quad ; \quad S = \dfrac{1}{\sqrt{2}} \left(s_r + i \, s_i \right) \, ,
\end{equation}
where $s_r (s_i)$ indicates the real (imaginary) components of the scalar singlet, $G^+ (G^0)$ represents the charged (neutral) Goldstone boson, while $h$ is the real part of the neutral component of the scalar doublet. The vacuum expectation values (vev) of the corresponding fields are given as
\begin{equation} \label{eq_tree_vacuum_config}
    \langle h \rangle = v_h \quad ; \quad \langle s_r \rangle = v_r \quad ; \quad \langle s_i \rangle = v_i \, .
\end{equation}
It is important to note that in the presence of non-self hermitian couplings in the scalar potential defined in Eq.~\ref{eq_Vtree}, the scalar singlet can obtain a complex vev, which is the most general vacuum configuration possible within this setup. The minimization conditions for the aforementioned vev configurations are given as
\begin{eqnarray}
    \mu_H^2 &=& \two \left[\roottwo v_r \delta_1 + v_i^2 (\delta_2 - \delta_3) + v_r^2 (\delta_2 + \delta_3) + \lambda v_h^2 \right] \, , \label{eq_tadpole1} \\
    b_1 &=& \dfrac{1}{24 v_r} \bigg[-24 \roottwo a + 6 \roottwo \mu_3 (v_i^2 - 3 v_r^2) - 2 \roottwo \widetilde{\mu}_3 (v_i^2 + v_r^2) - 3 \roottwo v_h^2 \delta_1 \nonumber \\
        &-& 6 v_r (4 d_1 (v_r^2 - v_i^2) + d_3 (v_r^2 + v_i^2) + 2 v_h^2 \delta_3) \bigg] \, , \label{eq_tadpole2} \\
    b_2 &=& \dfrac{1}{24 v_r} \bigg[-24 \roottwo a + 6 \roottwo \mu_3 (v_i^2 + v_r^2) - 2 \roottwo \widetilde{\mu}_3 (v_i^2 + 5 v_r^2) - 3 \roottwo v_h^2 \delta_1 \nonumber \\
    &+& 6 v_r ((2 d_1 - 2 d_2 + d_3) v_i^2 + (2 d_1 - 2 d_2 - d_3) v_r^2 - 2 v_h^2 \delta_2) \bigg] \ . \label{eq_tadpole3}
\end{eqnarray}

We now proceed to discuss the mass spectrum of the neutral scalars, which is an essential ingredient for the study of phase transitions. In the absence of well defined CP eigenstates, the neutral squared mass matrix $\mathcal{M}_0^2$ is a symmetric $3 \times 3$ Hermitian matrix that can be written in the $\left(h \quad s_r \quad s_i \right)^\mathsf{T}$ basis as 
\begin{equation} \label{eq_mass_matrix_neutral}
    \mathcal{M}_0^2 = 
    \begingroup 
    \setlength\arraycolsep{10pt}
    \begin{pmatrix}
        \two \lambda v_h^2 & \four v_h \left(\roottwo \delta_1 + 2 v_r (\delta_2 + \delta_3) \right) & \two v_h v_i (\delta_2 - \delta_3) \\[3ex]
        \cdot & \begin{aligned}
            \dfrac{1}{24 v_r} \Big[ &-24 \roottwo a_1 + 6 \roottwo c_{12} v_r^2 + 12 d_{123} v_r^3 \\[-0.4em]
            &+ \roottwo \left( (6 \mu_3 - 2 \widetilde{\mu}_3) v_i^2 - 3 v_h^2 \delta_1 \right) \Big]
        \end{aligned} & \begin{aligned}
            \six v_i \Big[ &\roottwo(-3 \mu_3 + \widetilde{\mu}_3)  \\[-0.4em]
            &+ 3 (-3 d_1 + d_2) v_r \Big]
        \end{aligned} \ \\[4ex]
        \cdot & \cdot & \two (d_1 + d_2 - d_3) v_i^2
    \end{pmatrix} \, ,
    \endgroup
\end{equation}
where $c_{12} = \mu_3 + \widetilde{\mu}_3$ and $d_{123} = d_1 + d_2 + d_3$. The squared mass matrix can be diagonalized by an orthogonal transformation to obtain the masses of the physical eigenstates ($H_1, H_2$, and $H_3$) through
\begin{equation}
    O \, \mathcal{M}_0^2 \, O^\mathsf{T} = \text{diag}\left(M_1^2, \ M_2^2, \ M_3^2 \right) \, ,
\end{equation}
where $M_i^2$ is the $ ({\rm mass})^2$ of $H_i$ for $i = 1, 2$ and $3$. The orthogonal matrix relates the gauge and mass eigenbasis by $\left(H_1 \quad H_2 \quad H_3 \right)^\mathsf{T} = O \left(h \quad s_r \quad s_i \right)^\mathsf{T}$ and is constructed as
\begin{equation}
    O = \begin{pmatrix}
            1 & 0 & 0 \\
            0 & c_1 & s_1 \\
            0 & -s_1 & c_1
        \end{pmatrix} \begin{pmatrix}
                          c_2 & 0 & s_2 \\
                          0 & 1 & 0 \\
                          -s_2 & 0 & c_2
                      \end{pmatrix} \begin{pmatrix}
                                        c_3 & s_3 & 0 \\
                                        -s_3 & c_3 & 0 \\
                                        0 & 0 & 1
                                    \end{pmatrix} \, ,
\end{equation}
with the short-hand notation for the three mixing angles given as $c_i \equiv \cos \theta_i$ and $s_i \equiv \sin \theta_i$ for $i = 1, 2$ and $3$.

Using the minimization conditions as given in Eqs.~\ref{eq_tadpole1}, \ref{eq_tadpole2}, and \ref{eq_tadpole3}, along with trading all the quartic couplings in terms of physical masses and mixing angles as given in Appendix \ref{app_lag_to_physical}, we are left with 13 input parameters for the numerical analysis
\begin{equation} \label{eq_inputs}
    \left\lbrace M_1, \ M_2, \ M_3, \ \theta_1, \ \theta_2, \ \theta_3, \ v_h, \ v_r, \ v_i, \ a, \ \delta_1, \ \mu_3, \ \widetilde{\mu}_3 \right\rbrace,
\end{equation}
where $v_h = 246$ GeV and $M_1 = 125$ GeV are kept fixed throughout the analysis, with the remaining 11 being free parameters.


\subsection{Constraints}

In this section, we discuss various theoretical and experimental constraints that were employed throughout the analysis. The theoretical constraints on the parameters of the scalar potential given in Eq. \ref{eq_Vtree} are now in order. \\

\noindent
\textbf{Unitarity :} Tree-level unitarity of different $2 \leftrightarrow 2$ scattering processes places constraints on the parameters of the scalar potential and has been extensively studied in the literature for the SM as well as for various extended scalar sector scenarios. Scatterings of massive gauge bosons can also be taken into account by replacing their longitudinal modes with their corresponding Goldstone bosons, utilizing the high energy limit of the equivalence theorem \cite{Lee:1977eg}. Moreover, scalar-scalar interactions are dominated by only their quartic interactions, and the analysis has been performed on the gauge basis as given in Eq. \ref{eq_field_components} due to the simpler form of the quartic couplings. The $2 \leftrightarrow 2$ scalar scattering processes leads to the $16 \times 16$ S-matrix, which can be further decomposed into five block diagonal submatrices consisting of three zero-charged channels $ \mathcal{M}_1 (4 \times 4)$, $\mathcal{M}_2 (5 \times 5)$, $\mathcal{M}_3 (2 \times 2)$, one singly-charged channel $\mathcal{M}_4 (4 \times 4)$ and a unique doubly-charged channel $\mathcal{M}_5 (1 \times 1)$.

The submatrix $\mathcal{M}_1$ corresponds to the scattering of the initial and final states of any of the states $\left\lbrace h s_i, G^0 s_r, G^0 s_i, h s_r \right\rbrace$ and is given by
\begin{equation}
    \mathcal{M}_1 = \begin{pmatrix}
        \two (\delta_2 - \delta_3) & 0 & 0 & 0\\
        0 & \two (\delta_2 + \delta_3) & 0 & 0\\
        0 & 0 & \two (\delta_2 - \delta_3) & 0\\
        0 & 0 & 0 & \two (\delta_2 + \delta_3)
    \end{pmatrix} \, ,
\end{equation}
with the eigenvalues given as
\begin{equation}
    \mathcal{E}_1 = \left\lbrace \two (\delta_2 - \delta_3), \two (\delta_2 + \delta_3) \right\rbrace \, .
\end{equation}

The $\mathcal{M}_2$ submatrix can be constructed similarly, with the initial and final states being one of the bases $\left\lbrace G^+ G^-, \roottwoinv G^0 G^0, \roottwoinv s_i s_i, \roottwoinv h h, \roottwoinv s_r s_r \right\rbrace$ and is given by
\begin{equation}
    \mathcal{M}_2 = \begin{pmatrix}
        \lambda & \roottwoinv \lambda & \roottwoinv (\delta_2 - \delta_3) & \roottwoinv \lambda & \roottwoinv (\delta_2 + \delta_3) \\
        \roottwoinv \lambda & 3 \lambda & \delta_2 - \delta_3 & \lambda & \delta_2 + \delta_3 \\
        \roottwoinv (\delta_2 - \delta_3) & \delta_2 - \delta_3 & 3 (d_1 + d_2 - d_3) & \delta_2 - \delta_3 & -3d_1 + d_2 \\
        \roottwoinv \lambda & \lambda & \delta_2 - \delta_3 & 3 \lambda & \delta_2 + \delta_3 \\
        \roottwoinv (\delta_2 + \delta_3) & \delta_2 + \delta_3 & -3d_1 + d_2 & \delta_2 + \delta_3 & 3 (d_1 + d_2 + d_3)
    \end{pmatrix} \, ,
\end{equation}
and the eigenvalues $\mathcal{E}_2$ have been found by numerically solving the characteristic polynomial of order five.

The third submatrix $\mathcal{M}_3$ can be constructed considering the basis $\left\lbrace h G^0, s_r s_i \right\rbrace$ which is given by
\begin{equation}
    \mathcal{M}_3 = \begin{pmatrix}
        \two \lambda & 0 \\
        0 & \two (-3 d_1 + d_2)
    \end{pmatrix} \, ,
\end{equation}
and the eigenvalues are given by
\begin{equation}
    \mathcal{E}_3 = \left\lbrace \two \lambda,  \two (-3 d_1 + d_2)\right\rbrace \, .
\end{equation}

The submatrix $\mathcal{M}_4$ consists of singly charged channels with ingoing and outgoing states as any of the following sets $\left\lbrace h G^+, s_r G^+, G^0 G^+, s_i G^+ \right\rbrace$ and is given by
\begin{equation}
    \mathcal{M}_4 = \begin{pmatrix}
        \two \lambda & 0 & 0 & 0\\
        0 & \two (\delta_2 + \delta_3) & 0 & 0\\
        0 & 0 & \two \lambda & 0\\
        0 & 0 & 0 & \two (\delta_2 - \delta_3)
    \end{pmatrix} \, ,
\end{equation}
with the eigenvalues as
\begin{equation}
    \mathcal{E}_4 = \left\lbrace \two \lambda, \two (\delta_2 + \delta_3), \two (\delta_2 - \delta_3) \right\rbrace \, .
\end{equation}

Finally, the submatrix of the unique doubly-charged channel with the ingoing and the outgoing state as $\roottwoinv G^+ G^+$ is given by
\begin{equation}
    \mathcal{M}_5 = \mathcal{E}_5 = 2 \lambda \, .
\end{equation}

To achieve unitarity, the absolute values of all the eigenvalues (all the components of $\mathcal{E}_1$, $\mathcal{E}_2$, $\mathcal{E}_3$, $\mathcal{E}_4$ and $\mathcal{E}_5$) of these submatrices should be less than $8 \pi$. \\

\noindent
\textbf{Vacuum stability :} Constraints also arise from ensuring that the scalar potential given in Eq. \ref{eq_Vtree} is bounded from below (BFB) from all field directions. The copositivity criteria \cite{Kannike:2012pe, Chakrabortty:2013mha} have been used to determine the BFB conditions. Taking into account only the quartic terms, the potential can be written in terms of charged ($G^+$) and neutral components ($\phi^0, S$) of the scalar fields and is given by
\begin{equation}
    \begin{aligned}
        V^{(4)}_0 (G^+, \phi^0, S) &= \four \bigg[\lambda \left({\phi^0}^{*2}{\phi^0}^2  + 2 \phi^0 {\phi^0}^* G^+ G^- + {G^+}^2 {G^-}^2 \right) \\
        &+ 2 \delta_3 |S|^2 G^+ G^- + (d_1 + d_2 + d_3) |S|^4 \bigg] \, ,
    \end{aligned}
\end{equation}
which can be expressed in the form of $V^{(4)}_0 = \Phi^\dagger \mathcal{A} \Phi$ where $\Phi = (G^+ \quad \phi^0 \quad S)^\mathsf{T}$ and $\mathcal{A}$ is known as the copositive matrix. We found that all the 2-field BFB conditions are subsets of the 3-field BFB conditions and are given by
\begin{equation}
    \begin{aligned}
        &\left\lbrace \lambda \geqslant 0, d_{123} \geqslant 0, \lambda d_{123} \geqslant 0, \delta_3 + \sqrt{\lambda d_{123}} \geqslant 0, \delta_3 + \sqrt{2 \lambda d_{123} + 2 \delta_3 \sqrt{\lambda d_{123}}} + \sqrt{\lambda d_{123}} \geqslant 0 \right\rbrace \, .
    \end{aligned}
\end{equation}

\noindent
\textbf{Perturbativity :} Demanding that the model remain perturbative across any scale imposes the following conditions
\begin{equation}
    |\lambda| < 4 \pi, |\delta_2 | < 4 \pi, |\delta_3| < 4 \pi, |d_1| < 4 \pi, |d_2| < 4 \pi, |d_3| < 4 \pi \, .
\end{equation}

Along with the theoretical constraints, there are additional relevant experimental results that severely constrain the BSM parameter space in various extended scalar scenarios, which are discussed next. \\

\noindent
\textbf{Electroweak precision tests :} The oblique electroweak parameters \cite{Peskin:1990zt, Grimus:2008nb}, namely the parameters $S, T$ and $U$, encode the corrections in the propagator of the EW gauge bosons, which can put stringent constraints on many extended scalar scenarios. In the presence of non-vanishing vevs of the singlet in cxSM, the singlet fields provide non-negligible quantum corrections to the self energies of $W^\pm$ and $Z$ bosons via mixing with the SM Higgs doublet. The parameters $S, T$, and $U$ can be expressed in terms of the physical masses and mixing angles of our model and are given by
\begin{align}
    S &= \dfrac{1}{24 \pi} \Big[ (c_1^2 c_2^2 - 1) \left( \log (M_1^2) + G(M_1^2, m_Z^2) \right) \nonumber \\
      &+ s_1^2 c_2^2 \left( \log (M_2^2) + G(M_2^2, m_Z^2) \right) + s_2^2 \left( \log (M_3^2) + G(M_3^2, m_Z^2) \right) \Big] \, , \\
    T &= \dfrac{3}{16 \pi s_W^2 m_W^2} \Big[ (c_1^2 c_2^2 - 1) \left( F(m_Z^2, M_1^2) - F(m_W^2, M_1^2) \right) \nonumber \\
      &+ s_1^2 c_2^2 \left( F(m_Z^2, M_2^2) - F(m_W^2, M_2^2) \right) + s_2^2 \left( F(m_Z^2, M_3^2) - F(m_W^2, M_3^2) \right) \Big] \, , \\
    U &= \dfrac{1}{24 \pi} \Big[ (c_1^2 c_2^2 - 1) \left( G(M_1^2, m_Z^2) - G(M_1^2, m_W^2) \right) \nonumber \\
      &+ s_1^2 c_2^2 \left( G(M_2^2, m_Z^2) - G(M_2^2, m_W^2) \right) + s_2^2 \left( G(M_3^2, m_Z^2) - G(M_3^2, m_W^2) \right) \Big] \, ,
\end{align}
where the functions $G(x, y)$ and $F(x, y)$ are defined as
\begin{align}
    G(x, y) &= -\dfrac{79}{3} + 9\dfrac{x}{y} - 2\dfrac{x^2}{y^2} + \left(-10 + 18 \dfrac{x}{y} - 6\dfrac{x^2}{y^2} + \dfrac{x^3}{y^3} - 9 \dfrac{x - y}{x+y} \right) \log \dfrac{x}{y} \nonumber \\
            &+ \dfrac{1}{y} \left( 12 - 4\dfrac{x}{y} + \dfrac{x^2}{y^2} \right) f(x, x^2 - 4 x y) \, , \\
    F(x, y) &= \left[ \dfrac{x + y}{2} - \dfrac{x y}{x - y} \log \dfrac{x}{y} \right] \left(1 - \delta_{x y} \right) \, , \\
    f(x, y) &= \sqrt{y} \log \left| \dfrac{x - \sqrt{y}}{x + \sqrt{y}} \right| \Theta(y) + 2 \sqrt{-y} \tan^{-1} \dfrac{\sqrt{-y}}{x} \Theta(-y) \, ,
\end{align}
with $\delta_{i j}$ and $\Theta$ being the Kronecker delta and the Heaviside theta function, respectively. The results of the fit to the electroweak precision data are \cite{Baak:2014ora}
\begin{equation}
    S \equiv \hat{S} + \sigma_S = 0.05 \pm 0.11; \, T \equiv \hat{T} + \sigma_T = 0.09 \pm 0.13; \, U \equiv \hat{U} + \sigma_U = 0.01 \pm 0.11 \, ,
\end{equation}
with the correlation coefficients as
\begin{equation}
    \rho_{ST} = 0.90; \, \rho_{TU} = -0.83; \, \rho_{SU} = -0.59 \, .
\end{equation}
We construct the test statistic $\chi^2 = X^\mathsf{T} \mathcal{C}^{-1} X$ with $X = \left( S - \hat{S} \quad T - \hat{T} \quad U - \hat{U} \right)^\mathsf{T}$ with the covariance matrix given as
\begin{equation}
    \mathcal{C} = \begin{pmatrix}
                      \sigma_S^2 & \rho_{S T} \sigma_S \sigma_T & \rho_{S U} \sigma_S \sigma_U \\
                      \rho_{S T} \sigma_S \sigma_T & \sigma_T^2 & \rho_{T U} \sigma_T \sigma_U \\
                      \rho_{S U} \sigma_S \sigma_U & \rho_{T U} \sigma_T \sigma_U & \sigma_U^2
                  \end{pmatrix} \, ,
\end{equation}
and put constraints on the parameter space at $95 \%$ confidence level utilizing $\chi^2 \leqslant 6.18$. \\

\noindent
\textbf{Higgs observables :} The mixing angles of the BSM scalars with Higgs, i.e, $\theta_2$ and $\theta_3,$ can be constrained from direct scalar searches and precision Higgs measurements at the LHC \cite{Lane:2024vur}. The scalar mixing angles considered in our analysis lie well within the experimental upper bound reported in \cite{Lewis:2024yvj, Robens:2025nev} for additional scalars in the absence of any $\mathbb{Z}_2$ symmetry.

All the aforementioned theoretical as well as experimental constraints were implemented during the numerical scan to find the parameter space resulting in SFOPT. We now proceed to discuss the construction of the effective potential for the cxSM model, which is the key ingredient for analyzing the dynamics of phase transitions.


\section{Phase transition } \label{sec_phase_transition}

In the presence of additional scalar fields, SFOPT can be achieved, unlike in the case of SM. In order to study the dynamics of cosmological phase transitions, the key ingredient is the effective thermal potential, which comprises of one-loop thermal corrections to the zero-temperature potential. In this section, we delve into the details of the effective potential and present the parameter space that exhibits SFOPT along with several benchmark points.


\subsection{Effective potential} \label{subsec_eff_pot}

The construction of a one-loop effective thermal potential contains quantum corrections at finite temperature, which can be decomposed into a zero-temperature correction known as the Coleman-Weinberg potential ($V_\text{CW}$) \cite{Coleman:1973jx} and a temperature dependent part denoted as $V_{T}$. The CW contribution to the tree level potential in the Landau gauge\footnote{The gauge independent analysis of effective potential remains beyond the scope of this work; for details, see Ref. \cite{Nielsen:1975fs,Fukuda:1975di} } is given by
\begin{equation} \label{eq_Vcw}
    V_{\rm CW} = \dfrac{1}{64 \pi^2} \sum_j (-1)^{F_j} n_j M_j^4 \left[ \log \dfrac{M_j^2 (H_i)}{\mu^2} - C_j - C_\epsilon \right]
\end{equation}
in the $\overline{\text{MS}}$ scheme where $M_j$ is the field dependent mass of the $j-$th particle with $n_j$ degrees of freedom. The constant $C_j = 3/2 \, (5/6)$ for scalars and fermions (gauge bosons), whereas the overall sign is determined by $F_j = 0$ and $F_j = 1$ for bosons and fermions, respectively. The renormalization scale has been set at the value of the top mass, i.e., $\mu \simeq 173$ GeV \footnote{The dependence of GW parameters on the renormalization scale has been extensively studied in Refs. \cite{Croon:2020cgk, Gould:2021oba} which we have briefly explored in Appendix \ref{app_mu_vary_on_gw} considering a particular benchmark point.}. The divergent part of the CW potential, denoted as $C_\epsilon = 2/\epsilon - \gamma_E + \log(4\pi)$ can be absorbed by introducing the counter terms given by
\begin{equation} \label{eq_Vct}
\begin{aligned}
    V_\text{CT} &= - \two \delta \mu_H^2 H^\dagger H + \four \delta\lambda \left(H^\dagger H \right)^2 + \left( \delta a S + \four \delta b_1 S^2 + \hc \right) + \two \delta b_2 |S|^2 \\
    &+ \left(\six \delta \mu_3 S^3 + \delta \widetilde{\mu}_3 |S|^2 S + \hc \right) + \left( \eight \delta d_1 S^4 + \eight \delta d_3 |S|^2 S^2 + \hc \right) + \four \delta d_2 |S|^4 \\
    &+ H^\dagger H  \left[ \left(\four \delta \delta_1 S + \four \delta \delta_3 S^2 + \hc \right) + \two \delta \delta_2 |S|^2 \right]
\end{aligned} \, .
\end{equation}
where the coefficients, which are explicitly given in Appendix \ref{app_counter_terms}, are fixed by the renormalization conditions given by
\begin{equation} \label{eq_renorm_cond}
    \dfrac{\partial \left(V_\text{CW} + V_\text{CT} \right) }{\partial x} = 0  \, \quad ; \quad  \dfrac{\partial^2 \left(V_\text{CW} + V_\text{CT} \right) }{\partial x \partial y} = 0  \, ,
\end{equation}
evaluated at the vacuum configuration given in Eq.~\ref{eq_tree_vacuum_config}, where $x, y \in \lbrace h, s_r, s_i \rbrace$. This ensures that the tree level masses and vevs of the scalars remain unaltered. 

The one loop temperature dependent part of the effective potential is given by \cite{Dolan:1973qd, Quiros:1999jp}
\begin{equation} \label{finiteT}
    V_{T} (H_i, T) = \dfrac{T^4}{2\pi^2} \left[ \sum_B n_B J_B \left( \dfrac{M_B^2 (H_i)}{T^2} \right) + \sum_{F} n_F J_F \left( \dfrac{M_F^2 (H_i)}{T^2} \right) \right]
\end{equation}
where the bosonic and fermionic functions are denoted by $J_B(x)$ and $J_F(x)$ respectively and are given by 
\begin{equation}
    J_{B/F}(x^2) = \int_{0}^{\infty} k^2 \log \left[1 \mp e^{-\sqrt{x^2 + k^2}} \right]dk
\end{equation}
where the negative (positive) sign appears for the bosons (fermions). In the high temperature regime ($T \gg M_i$), these functions can be expanded in the following form given by 
 \begin{align}  \label{JbJf}
     J_{\rm B}(x^2) &= -\frac{\pi^4}{45} + \frac{\pi^2}{12} x^2 -\frac{\pi}{6}x^3 + {\cal O}(x^4) \, , \nonumber \\
     J_{\rm F}(x^2) &= \frac{7\pi^4}{360} - \frac{\pi^2}{24} x^2  + {\cal O}(x^4) \, .
 \end{align}
with $x \equiv M_i/T$. It should be noted that bosonic function suffer from infrared divergences arising from Matsubara-zero modes \cite{Matsubara:1955ws}, which can be taken care of by their corresponding thermal masses via daisy resummation. We implement this method in the Arnold-Espinosa scheme \cite{Arnold:1992rz} by considering the following additional daisy contributions to the effective potential given by
\begin{equation} \label{Vring}
    V_{\rm ring}(H_{i},T)=-\sum_{j}\frac{n_{j}T}{12\pi}\Big[\left(M_{j}^2(H_{i}) + \Pi_j \, T^2 \right)^{3/2}-M_{j}^3(H_{i})\Big] \, ,
\end{equation}
where $\Pi$ is the diagonal matrix containing all the daisy coefficients, which can be read off from the high-temperature expansion of $V_{T}(H_i, T)$ and are given by
\begin{equation}
   \begin{bmatrix}
    \dfrac{2\delta_2 + 6\lambda}{48}+\dfrac{3g_2^2+g_1^2}{16}+\dfrac{3(y_b^2+y_t^2) + y_\tau^2}{12} &0&0\\
    0& \dfrac{4 d_2 + 3 d_3 + 4(\delta_2+\delta_3)}{48} &0 \\0&0&\dfrac{4 d_2 - 3 d_3 + 4(\delta_2-\delta_3)}{48}
    \end{bmatrix}.
\end{equation}

Thus, combining all the contributions discussed above, the total effective potential is given by
 \begin{equation}
 \label{vtot}
     V_\text{eff} (H_i, T) = V_0 (H_i) + V_\text{CW} (H_{i}) + V_\text{CT} + V_{T}(H_{i}, T) + V_\text{ring}(H_{i},T) \, .
 \end{equation}

The critical temperature ($T_C$) of a phase transition is defined as the temperature at which the symmetric phase and the broken phase coexist, and it can be obtained by equating the potential at the field values of these two different phases \cite{Patel:2011th,Patel:2012pi}. The strength of an FOPT is determined by the parameter $\gamma_C$, which in our case is defined as
\begin{equation} \label{eq_gammac}
    \gamma_C = \sqrt{\sum_{i = 1}^3 \gamma_{C i}^2} = \sqrt{\sum_{i = 1}^3 \left( \dfrac{\Delta H_i}{T_C} \right)^2 } \ ,
\end{equation}
where $\gamma_{C i}$ is the individual strength defined as the ratio of the difference in vevs of the scalar field ($\Delta H_{i}$) and the critical temperature $T_C$. The criterion \cite{Quiros:1999jp} for SFOPT is chosen as $\gamma_C \geq 1$.

In cosmological phase transitions, in addition to the critical temperature, another temperature which is of significant importance is the bubble nucleation temperature ($T_N$), which is generally lower than $T_C$. The transition proceeds through the nucleation of bubbles of the true vacuum within the false vacuum. The probability of tunneling from the false to the true vacuum per unit volume at a finite temperature $T$ is given by \cite{Linde:1981zj},
\begin{equation}
    \Gamma(T) = T^4 \left(\frac{S_{3}}{2\pi T}\right)^{3/2} e^{-S_{3}/T} \, ,
\end{equation}
where $S_3$ is the three-dimensional Euclidean action evaluated for the bounce solution and is given by

\begin{equation}
\label{3action}
    S_{3}=4\pi \int  r^2 dr\left[\frac{1}{2}\left(\frac{d H_{i}}{dr}\right)^2 + V_\text{eff} (H_{i},T)\right] \, ,
\end{equation}
and the field $H_i$ follow the Euclidean equation of motion \cite{Enqvist:1991xw, Affleck:1980ac} given by 
\begin{equation}
    \frac{d^2H_{i}}{dr^2} + \frac{2}{r}\frac{dH_{i}}{dr}=\frac{\partial V_\text{eff} (H_{i},T)}{\partial H_{i}} \, , 
\end{equation}
with the boundary conditions $d H_i/dr=0$ at $r=0$ and $H_i = H_i^\infty$ as $r \rightarrow \infty$ where $H_i^\infty$ is the value of the $i$-th field in the false vacuum. The nucleation temperature can be obtained by satisfying the following condition \cite{Enqvist:1991xw}
\begin{equation} \label{nuctemp}
    \frac{S_{3}(T_N)}{T_N} \simeq 140 \, .
\end{equation}
In addition to the strength of the transition defined in terms of the critical temperature, one can also define the strength of the transition in terms of the nucleation temperature similar to Eq. \ref{eq_gammac} and is given by
\begin{equation} \label{eq_gamman}
    \gamma_N = \sqrt{\sum_{i = 1}^3 \gamma_{N i}^2} = \sqrt{\sum_{i = 1}^3 \left( \dfrac{\Delta H_i}{T_N} \right)^2 } \ ,
\end{equation}
where the difference in the vevs of the scalar fields are evaluated at $T_N$. We now present our parameter space scan and study of the various benchmark points listed in Table \ref{tab_bp_details}.


\subsection{Scan based analysis} \label{subsec_scan_based_analysis}

\newcommand{\scanRange}{
\begin{table}[t]
    \centering
    \renewcommand{\arraystretch}{1.25}
\resizebox{!}{!}{
\begin{tabular}{|c|c||c|c||c|c|}
\hline
Parameter & Range &  Parameter & Range & Parameter & Range \tabularnewline
\hline
\hline
$M_2$ [GeV] & [150, 300]  & $v_r$ [GeV] & [100, 200] & $\delta_1$ [GeV] & [-40, 40] \tabularnewline
\hline
$M_3$ [GeV] & [150, 300] &  $v_i$ [GeV] & [50, 150] & $\mu_3$ [GeV] & [-40, 40] \tabularnewline
\hline
$\sin \theta_\alpha$ & [-0.2, 0.2]  & $a$ [GeV$^3$] & [-100, 100] & $\widetilde{\mu}_3$ [GeV] & [-40, 40] \tabularnewline
\hline
\end{tabular}
}
\renewcommand{\arraystretch}{1.0}
\caption{Ranges of the parameters for the numerical scan where $\alpha = 1,2,3$.}
\label{tab_param_range}
\end{table}
}

\newcommand{\bpDetails}{
\begin{table}[t]
\centering
\renewcommand{\arraystretch}{1.75}
\resizebox{\textwidth}{!}{
\begin{tabular}{|c|c|c|c|c|c|c|c|c|c|c|c|c|}
\hline
 \multirow{2}{*}{} &  & $M_2$ & $M_3$ & \multirow{2}{*}{$\sin \theta_1$} & \multirow{2}{*}{$\sin \theta_2$} & \multirow{2}{*}{$\sin \theta_3$} & $v_r$ & $v_i$ & $a$ & $\delta_1$ & $\mu_3$ & $\widetilde{\mu}_3$\tabularnewline
 \cline{3-4} \cline{8-13}
 &  & [GeV] & [GeV] &  &  &  & [GeV] & [GeV] & [GeV$^3$] & [GeV] & [GeV] & [GeV]\tabularnewline
\hline
\hline
BP1& \textcolor[HTML]{d62728}{\ding{108}}& 220.225 & 290.731 & $1.608 \times 10^{-1}$ & $-4.779 \times 10^{-3}$ & $-7.376 \times 10^{-2}$ & 189.554 & 111.256 & 38.427 & -19.312 & 19.744 & -7.691\tabularnewline
\hline
BP2 &\textcolor[HTML]{2ca02c}{\ding{115}}& 198.053 & 279.991 & $-1.744 \times 10^{-1}$ & $5.288 \times 10^{-2}$ & $-1.158 \times 10^{-1}$ & 179.670 & 107.939 & 3.460 & 14.754 & 4.849 & -3.180\tabularnewline
\hline
BP3 &\textcolor[HTML]{1f77b4}{\ding{58}} & 204.300 & 269.592 & $-2.626 \times 10^{-2}$ & $1.133 \times 10^{-1}$ & $-1.183 \times 10^{-1}$ & 147.990 & 91.482 & 61.965 & 16.280 & 11.781 & -11.181\tabularnewline
\hline
BP4 &\textcolor[HTML]{ff7f0e}{\ding{116}}& 206.234 & 264.191 & $-8.856 \times 10^{-2}$ & $5.183 \times 10^{-5}$ & $8.578 \times 10^{-2}$ & 184.632 & 101.817 & 89.160 & -24.132 & 32.857 & -17.904\tabularnewline
\hline
BP5 &\textcolor[HTML]{9467bd}{\ding{72}}& 207.945 & 275.466 & $-1.044 \times 10^{-1}$ & $8.611 \times 10^{-2}$ & $1.115 \times 10^{-1}$ & 167.078 & 98.955 & 77.440 & 4.587 & 18.881 & -21.893\tabularnewline
\hline
BP6 &\textcolor[HTML]{8c564b}{\ding{54}}& 183.971 & 232.786 & $1.060 \times 10^{-1}$ & $6.085 \times 10^{-2}$ & $-1.588 \times 10^{-2}$ & 194.042 & 86.067 & -78.745 & -29.185 & 30.787 & -18.232\tabularnewline
\hline
BP7 &\textcolor[HTML]{bcbd22}{\ding{117}}& 183.658 & 267.481 & $-5.518 \times 10^{-2}$ & $-5.337 \times 10^{-2}$ & $-1.232 \times 10^{-1}$ & 191.677 & 105.541 & -18.386 & 21.760 & 7.529 & -14.066\tabularnewline
\hline
\end{tabular}
}
\renewcommand{\arraystretch}{1.0}
\caption{Numerical values of all the input parameters as listed in Eq.~\ref{eq_inputs} for the benchmark points exhibiting SFOPT. The colored marker in the second column refers to the corresponding benchmark point shown in Fig.~\ref{fig_param_space}.}
\label{tab_bp_details}
\end{table}
}

\scanRange

We implement the cxSM model in \texttt{PhaseTracer2} \cite{Athron:2024xrh, Athron:2020sbe} to find the latent heat released ($\alpha$) during SFOPT, the inverse duration ($\beta$), and the nucleation temperature $(T_N)$ of the transition, which are essential quantities for the generation of GW as discussed later. We choose a hierarchical scenario for the BSM scalar masses ($M_3 > M_2$) along with their vevs at $\mathcal{O}(100)$ GeV with other dimension full parameters below the electroweak scale. We perform an extensive numerical scan over the 11 input parameters given in Eq.~\ref{eq_inputs}, within the ranges specified in Table~\ref{tab_param_range}. We randomly generate $1.5 \times 10^5$ parameter points using uniform priors for each input parameter. Imposing perturbativity constraints retains about $\sim 76\%$ of the original parameter space, while the subsequent application of unitarity constraints further reduces it to $\sim 71\%$. Most of the remaining parameter space is allowed by vacuum stability and electroweak precision observables, leaving 106950 parameter points that satisfy all theoretical and experimental constraints. We then examine these points to identify regions exhibiting a SFOPT. We find that only a tiny fraction, about $\sim 0.6\%$ of the theoretically allowed parameter space, yields SFOPT.

\begin{figure}[t]
    \centering
    \begin{subfigure}{0.495\textwidth}
        \centering
        \includegraphics[width=1\linewidth]{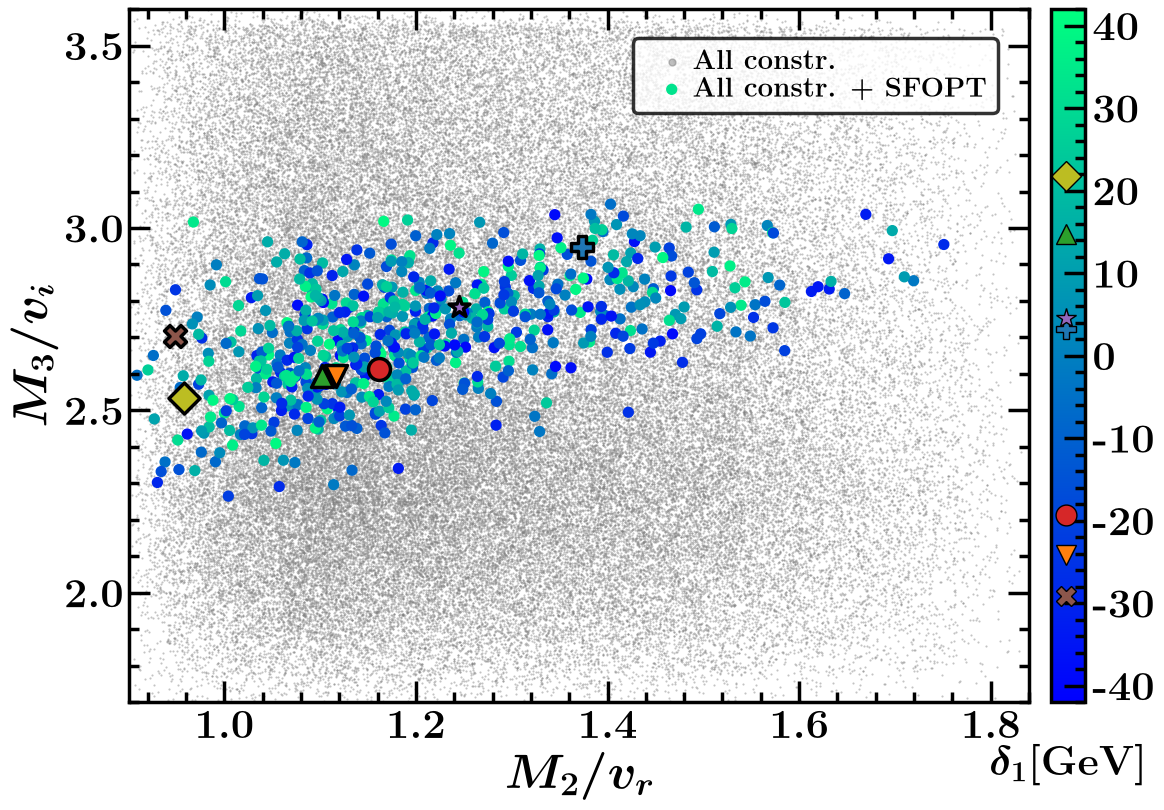}
        \caption{}
        \label{fig_param_space_mass_plane}
    \end{subfigure} \hfill
    \begin{subfigure}{0.495\textwidth}
        \centering
        \includegraphics[width=1\linewidth]{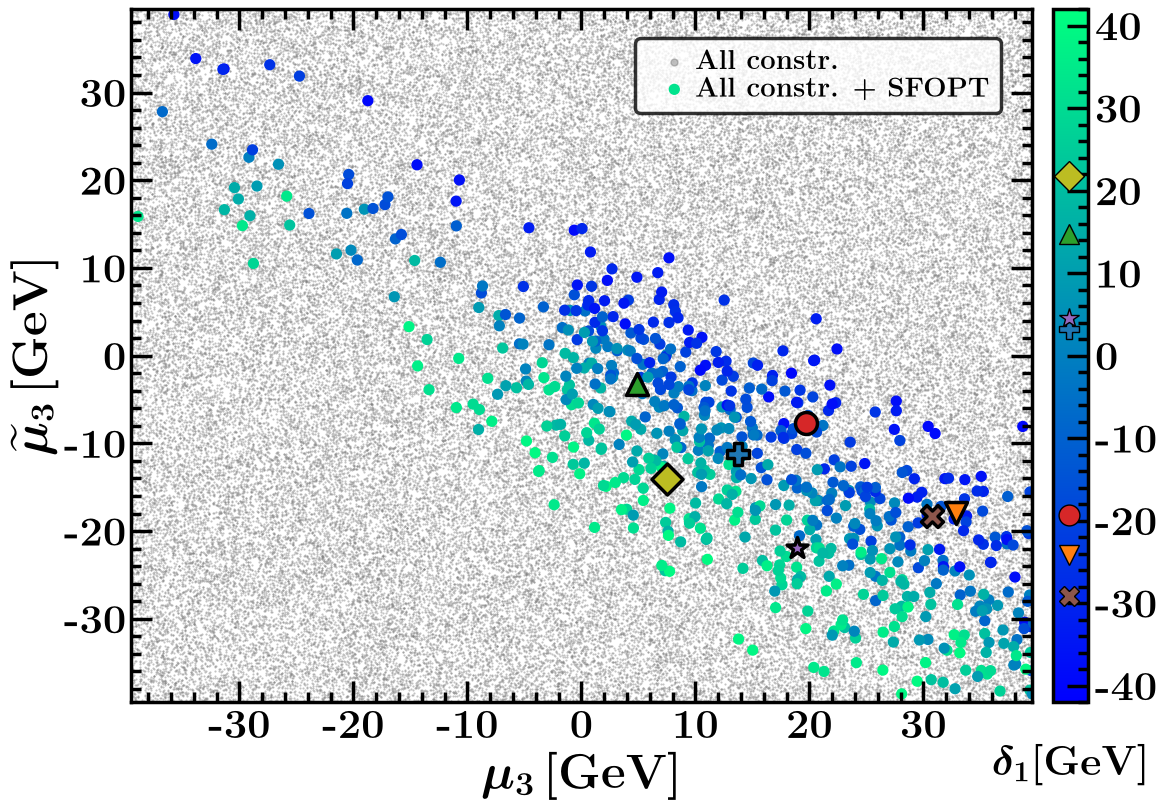}
        \caption{}
        \label{fig_param_space_c_plane}
    \end{subfigure}
    \caption{Parameter space satisfying SFOPT in the cxSM. The gray points are allowed by all the constraints, while the colored points exhibit SFOPT. The colored markers denote the benchmark points for which the details are presented in Table \ref{tab_bp_details}.}
    \label{fig_param_space}
\end{figure}

We present the parameter space satisfying all theoretical and experimental constraints in Fig.~\ref{fig_param_space} where the colored points satisfy SFOPT. Although the mixing angles were uniformly distributed in the specified ranges as mentioned in Table \ref{tab_param_range}, we observe that the region satisfying SFOPT lies within the range of $M_2/v_r \in [0.9, 1.7]$ and $M_3/v_i \in [2.25, 3.0]$ along with a negative correlation between the self cubic couplings $\mu_3$ and $\widetilde{\mu}_3$. We find single-step and multistep SFOPT along different field directions, which are categorized into three major types, which we proceed to discuss via several benchmark points (denoted by colored \ding{54} in Fig. \ref{fig_param_space} and shown in Table \ref{tab_bp_details}) in the following section.

\bpDetails


\subsection{Types}

\newcommand{\bpOutput}{
\begin{table}[t]
\centering
\renewcommand{\arraystretch}{1.25}
\resizebox{!}{!}{
\begin{tabular}{|c|c|c|c|c|c|c|c|c|}
\hline
 &  &  & $T_C$ [GeV] & $\gamma_C$ & $T_N$ [GeV] & $\gamma_N$ & $\alpha/10^{-3}$ & $\beta/H_N$\tabularnewline
\hline\hline
BP1 & \textcolor[HTML]{d62728}{\ding{108}} & I & 151.593 & 1.317 & 139.498 & 1.549 & 5.412 & 2861.602\tabularnewline
\hline
\multirow{2}{*}{BP2} & \multirow{2}{*}{\textcolor[HTML]{2ca02c}{\ding{115}}} & I & 169.232 & 1.259 & 156.956 & 1.321 & 1.656 & 4088.654\tabularnewline
\cline{3-9}
 &  & II & 150.843 & 0.926 & 146.301 & 1.070 & 1.862 & 6728.672\tabularnewline
\hline
\multirow{2}{*}{BP3} & \multirow{2}{*}{\textcolor[HTML]{1f77b4}{\ding{58}}} & I & 163.703 & 1.170 & 149.986 & 1.272 & 1.952 & 3159.387\tabularnewline
\cline{3-9}
 &  & II & 149.586 & 0.974 & 144.535 & 1.125 & 2.108 & 5927.351\tabularnewline
\hline
BP4 & \textcolor[HTML]{ff7f0e}{\ding{116}} & I & 134.191 & 1.567 & 120.798 & 1.910 & 9.348 & 2008.619\tabularnewline
\hline
BP5 & \textcolor[HTML]{9467bd}{\ding{72}} & I & 119.027 & 2.064 & 83.849 & 3.285 & 48.319 & 528.122\tabularnewline
\hline
BP6 & \textcolor[HTML]{8c564b}{\ding{54}} & I & 137.017 & 1.691 & 117.730 & 2.169 & 11.840 & 1283.953\tabularnewline
\hline
BP7 & \textcolor[HTML]{bcbd22}{\ding{117}} & I & 145.081 & 1.585 & 126.023 & 2.007 & 9.362 & 1466.208\tabularnewline
\hline
\end{tabular}
}
\renewcommand{\arraystretch}{1.0}
\caption{Relevant parameters of phase transition for the benchmark points of Table \ref{tab_bp_details}. The third column refers to a particular step in the case of a multistep phase transition.}
\label{tab_bp_output}
\end{table}
}

\newcommand{\bpOneWithVevPhase}{
\begin{figure}[ht]
    \centering
    \begin{subfigure}{0.5\linewidth}
        \centering
        \includegraphics[width=\linewidth]{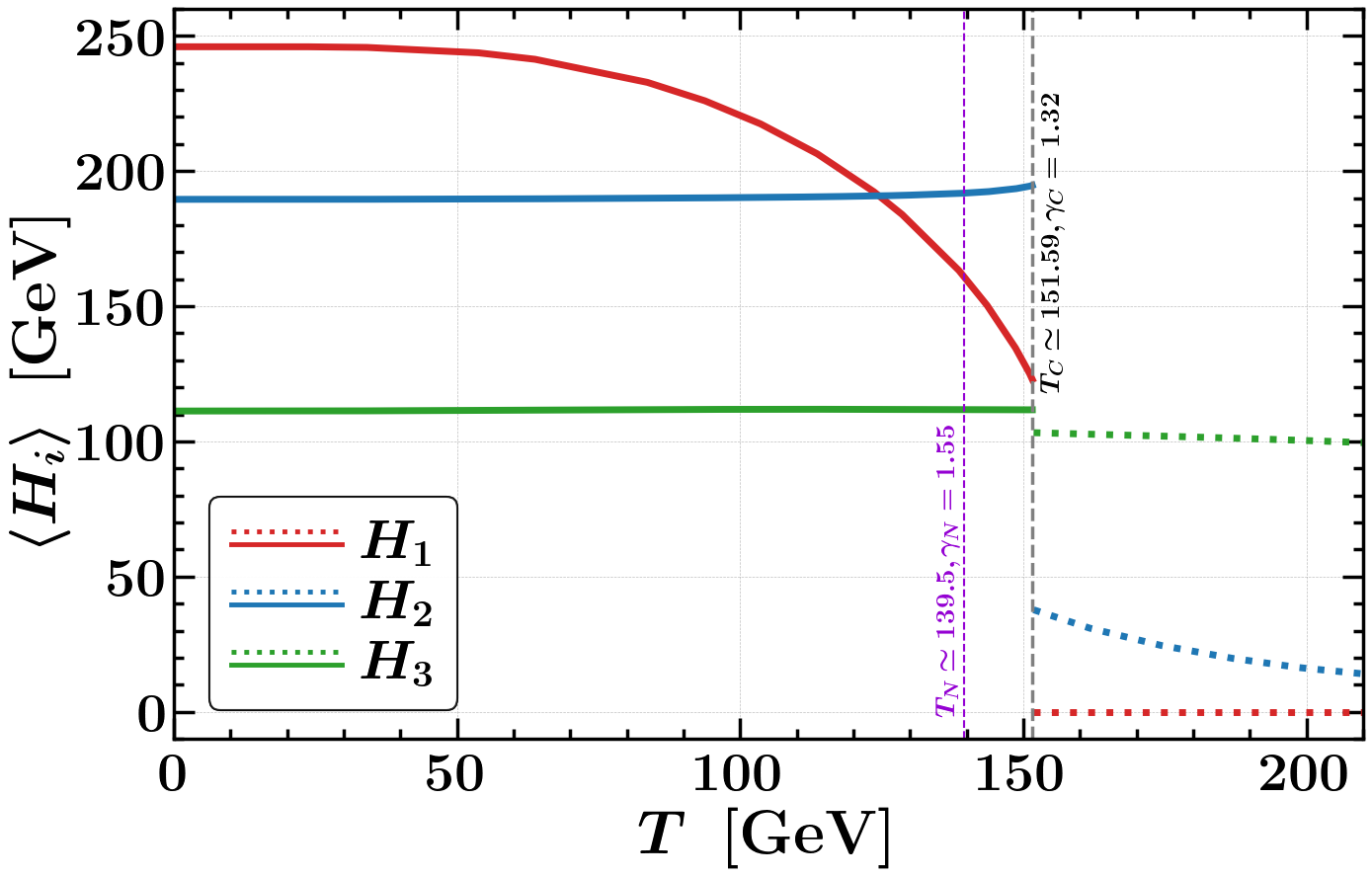}
        \caption{}
        \label{fig_bp1_vev_evolve}
    \end{subfigure} \hfill
    \begin{subfigure}{1\linewidth}
        \centering
        \includegraphics[width=0.95\linewidth]{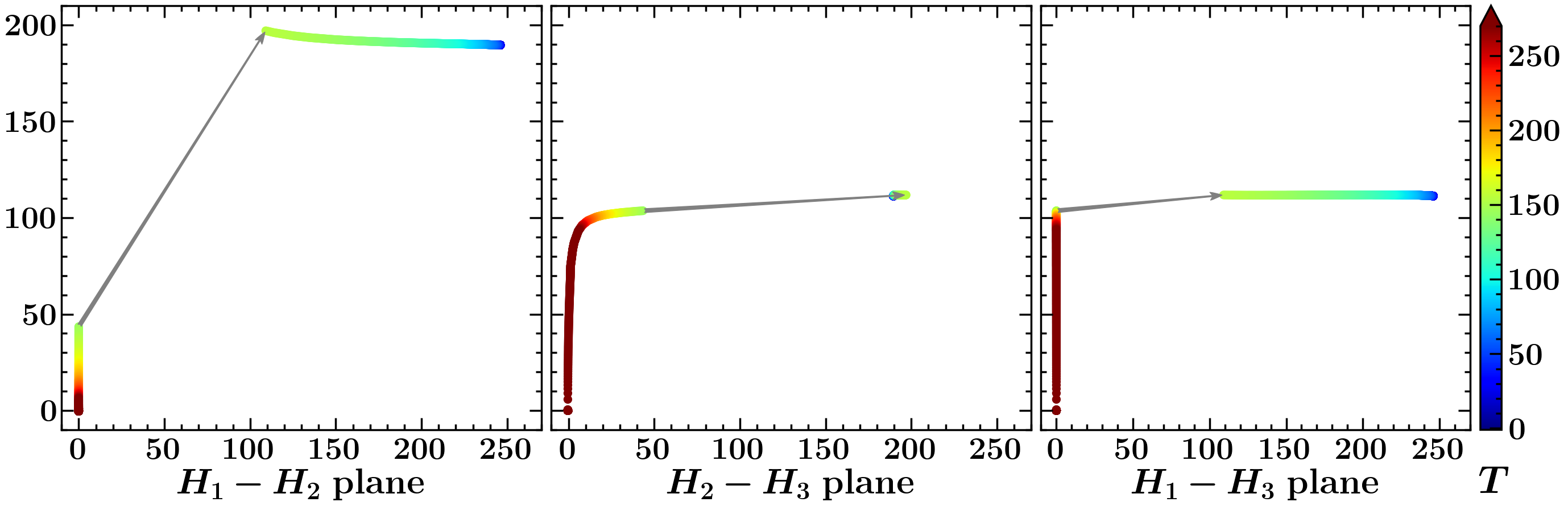}
        \caption{}
        \label{fig_bp1_phase_diagram}
    \end{subfigure}
    \caption{Upper panel (\ref{fig_bp1_vev_evolve}) shows the evolution of all the fields as a function of temperature for BP1, where the different phases are denoted via solid and dashed lines. The vertical dashed gray and purple lines denote the critical and nucleation temperatures, respectively. Lower panel (\ref{fig_bp1_phase_diagram}) depicts the phase diagram for BP1 in the three different field spaces, while temperature has been shown parametrically in the color bar with the arrows denoting the discontinuity of fields at the critical temperature. The numerical values of all the fields and temperature shown in the figures are in the units of GeV.}
\end{figure}
}

\newcommand{\bpTwoThreeVev}{
\begin{figure}[t]
    \centering
    \begin{subfigure}{0.49\linewidth}
        \centering
        \includegraphics[width=1\textwidth]{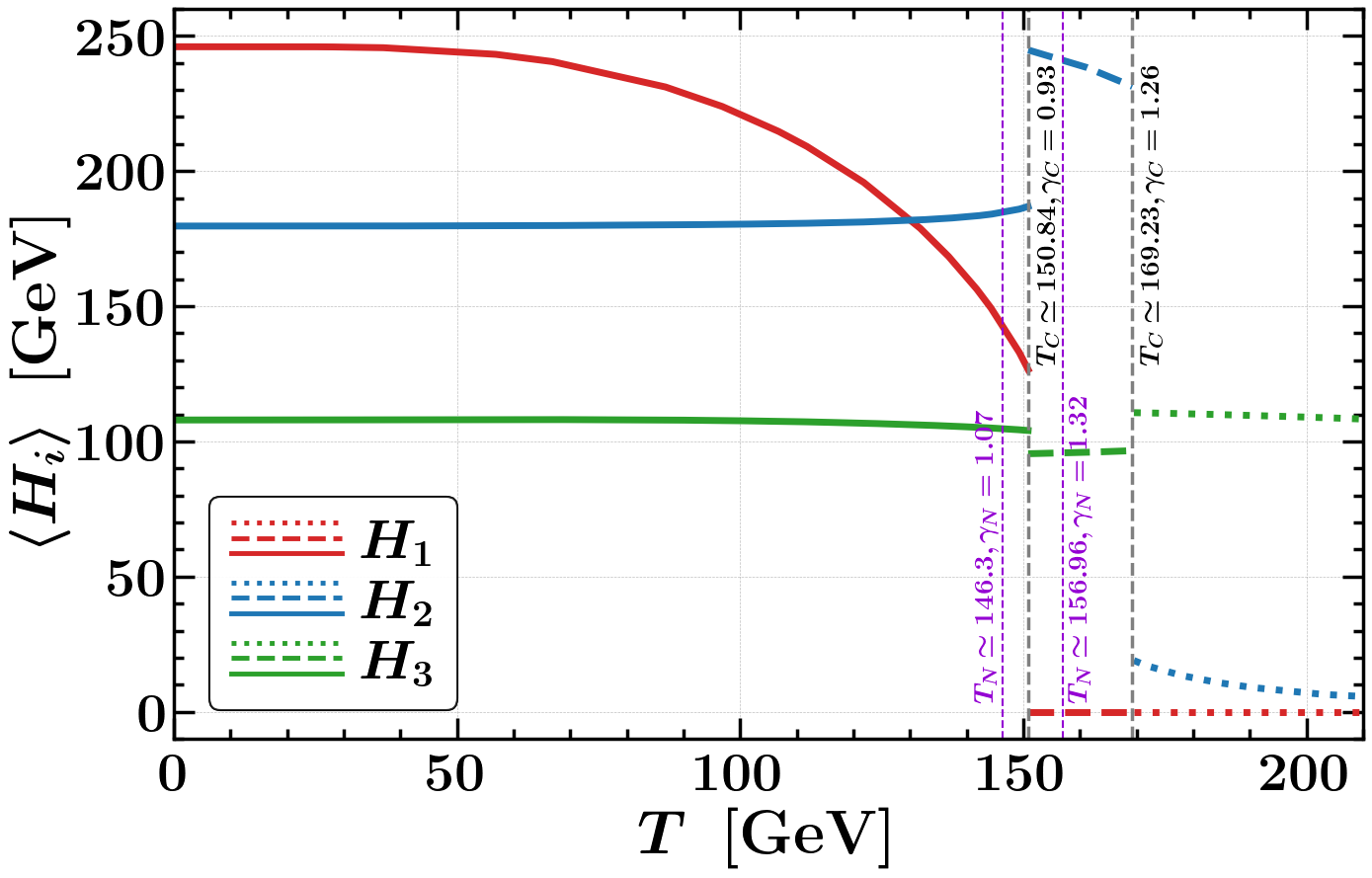}
        \caption{}
        \label{fig_bp2_vev_evolve}
    \end{subfigure} \hfill
    \begin{subfigure}{0.49\linewidth}
        \centering
        \includegraphics[width=1\textwidth]{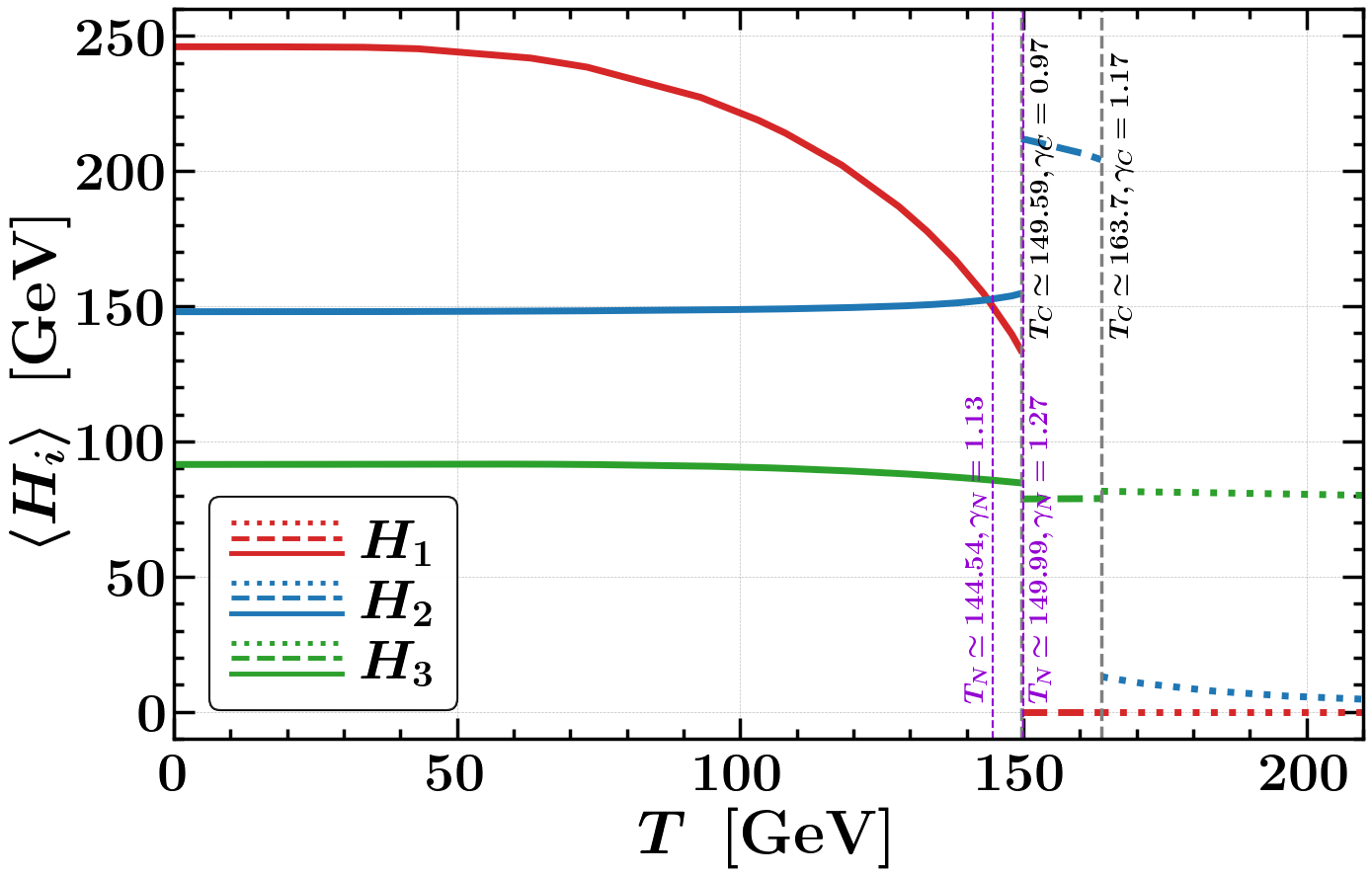}
        \caption{}
        \label{fig_bp3_vev_evolve}
    \end{subfigure}
    \caption{Same as Fig. \ref{fig_bp1_vev_evolve} for BP2 (left panel) and BP3 (right panel).}
\end{figure}
}

\newcommand{\bpFourFiveVev}{
\begin{figure}[t]
    \centering
    \begin{subfigure}{0.49\linewidth}
        \centering
        \includegraphics[width=1\textwidth]{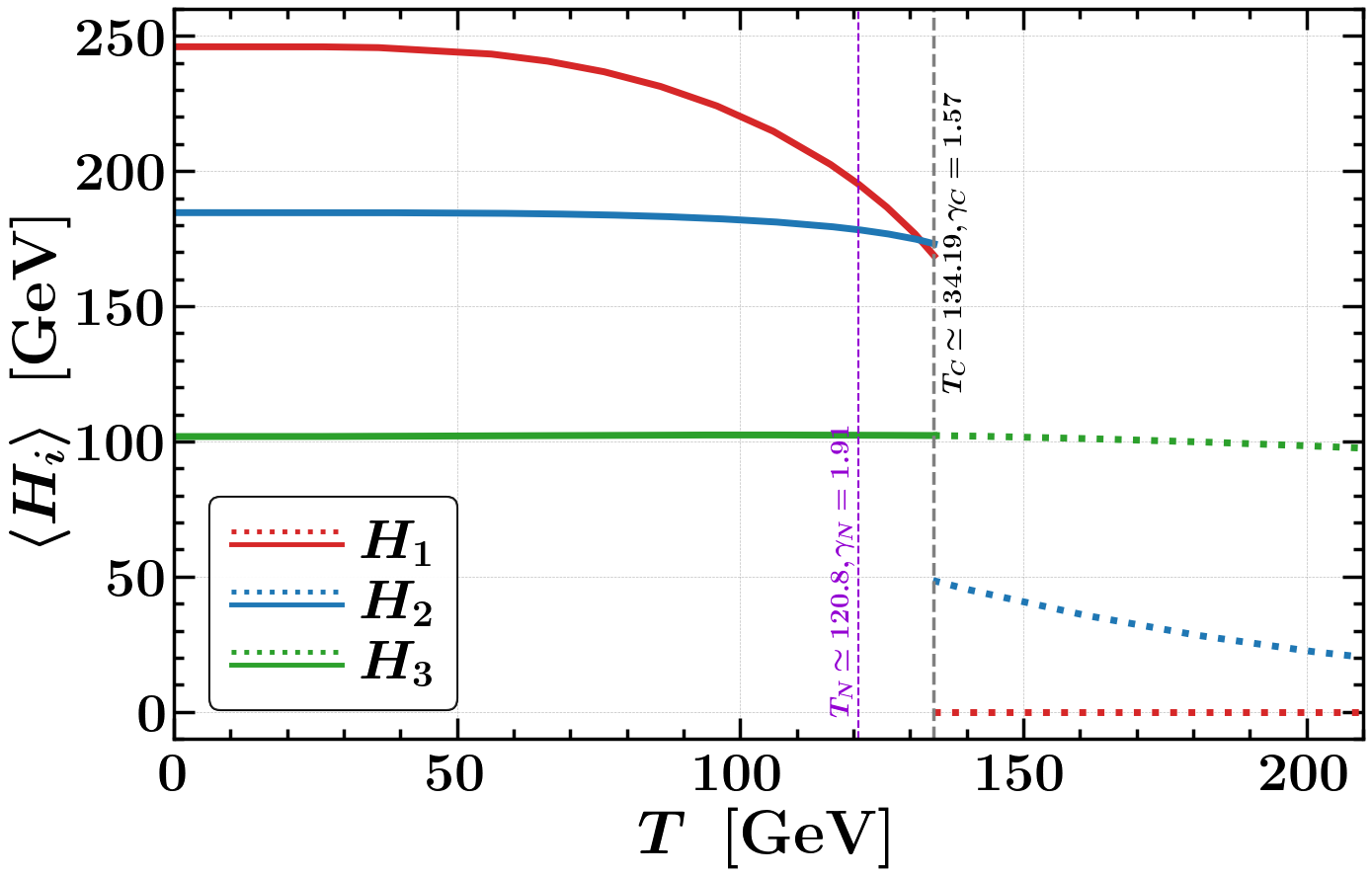}
        \caption{}
        \label{fig_bp4_vev_evolve}
    \end{subfigure} \hfill
    \begin{subfigure}{0.49\linewidth}
        \centering
        \includegraphics[width=1\textwidth]{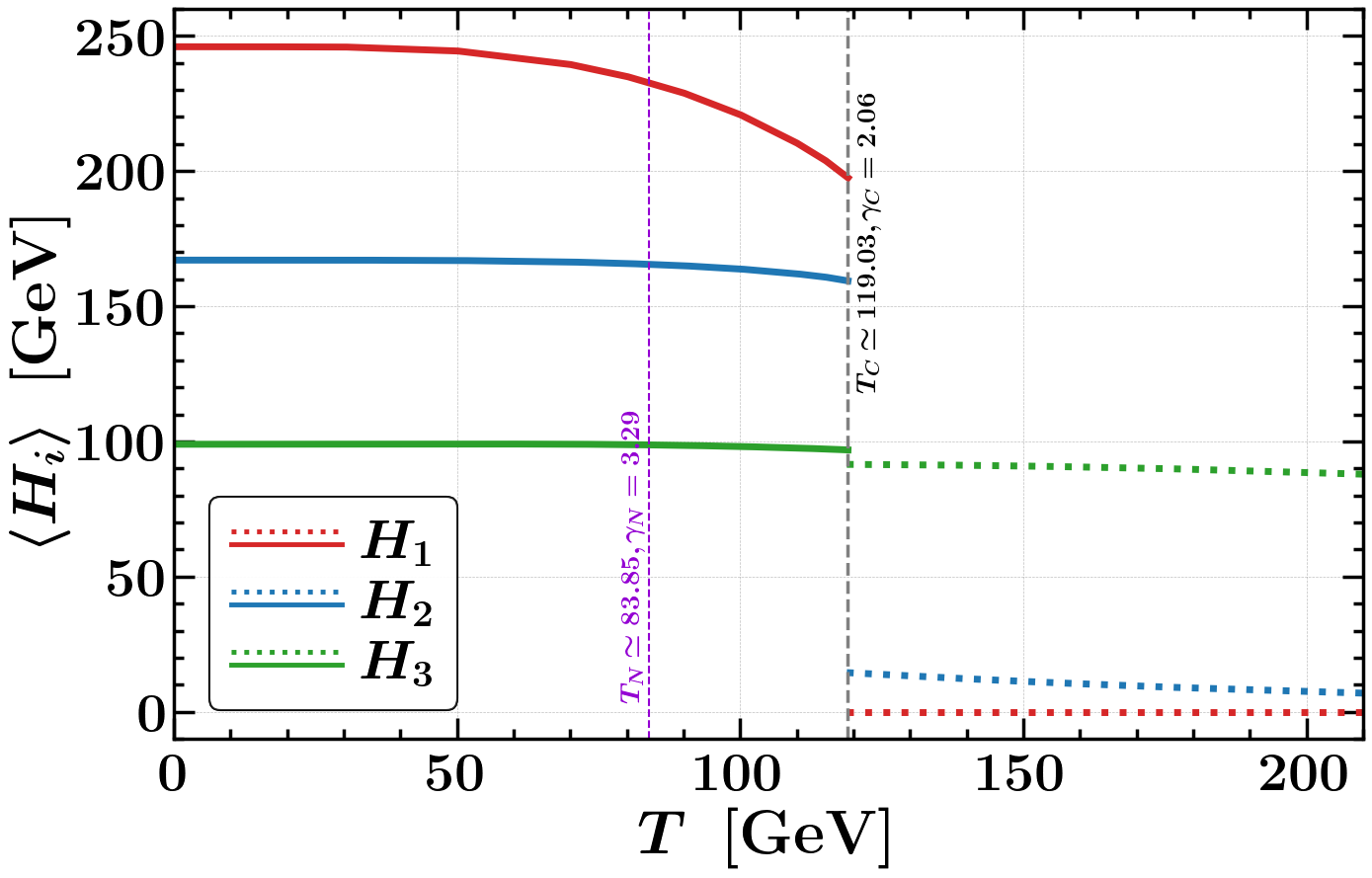}
        \caption{}
        \label{fig_bp5_vev_evolve}
    \end{subfigure}
    \caption{Same as Fig. \ref{fig_bp1_vev_evolve} for BP4 (left panel) and BP5 (right panel).}
\end{figure}
}

\bpOutput

From our numerical analysis, we analyze the dynamics of SFOPT via several benchmark points and classify them in three categories, where SFOPT occurs in the direction of the BSM (along $H_2$ and/or $H_3$ only), the SM (along $H_1$ only), and both SM + BSM (along $H_1$ and $H_2/H_3$).

\bpOneWithVevPhase

We first discuss the benchmark points where the SFOPT occurs only in the direction of BSM scalars, i.e, along the $H_2$ or $H_3$ direction. As can be seen from the phase structure for the first benchmark point BP1 shown in Fig.~\ref{fig_bp1_vev_evolve}, there is an SFOPT only along the $H_2$ direction at $T_C = 151.59$ GeV. This can be explicitly verified from the change in the fields given by $\Delta H_1 \simeq 123$ GeV, $\Delta H_2 \simeq 157$ GeV, $\Delta H_3 \simeq 8$ GeV resulting to individual strengths of $\gamma_{C 1} = 0.79, \gamma_{C 2} = 1.04, \gamma_{C 3} = 0.06$ respectively, leading to a combined strength of $\gamma_C = 1.32$ and $\gamma_N = 1.55$. The discontinuity of the field values is associated with a single step first order transition at the critical temperature, denoted by the vertical dashed gray line distinguishing the boundary between the symmetric phase and the broken phase. The bubbles of true vacuum start to nucleate at slightly lower temperature with $T_N = 139.5$ GeV as indicated by the vertical purple dashed line in Fig.~\ref{fig_bp1_vev_evolve}. The phase diagram has been represented in the $H_1 - H_2, H_2 - H_3$ and $H_1 - H_3$ space with the arrow explicitly denoting the discontinuity of the fields, as shown in Fig.~\ref{fig_bp1_phase_diagram}. In contrast to BP1, there can be phase transitions realized in multiple steps, of which the first step exhibits SFOPT, as shown in Figs.~\ref{fig_bp2_vev_evolve} and \ref{fig_bp3_vev_evolve} for BP2 and BP3, respectively. The benchmark point BP2 (BP3) exhibits an SFOPT in the first step along the $H_2$ direction at $T_C = 169.23 \ (163.7)$ GeV with $\gamma_C = 1.26 \ (1.17)$, whereas the second transition is a weak FOPT along all field directions.

\bpTwoThreeVev
\bpFourFiveVev

Next, we identify the benchmark point BP4 where a single step SFOPT occurs only in the direction of SM Higgs, i.e., $\gamma_{C 1} > 1$, as shown in Fig.~\ref{fig_bp4_vev_evolve}. The critical temperature is found to be $T_C = 134.19$ GeV with a total strength of $\gamma_C = 1.57$, whereas the nucleation temperature is found to be $T_N = 120.8$ GeV with $\gamma_N = 1.91$. The fields $H_1$ and $H_2$ exhibit a discontinuity at the critical temperature, whereas the field $H_3$ is continuous throughout the false and true vacuum.

Finally, we discuss the SFOPT achieved along both the SM and the BSM field directions. To illustrate this, we show the evolution of the fields for the benchmark point BP5 in Fig.~\ref{fig_bp5_vev_evolve} where it is evident that both $\gamma_{C 1}$ and $\gamma_{C 2}$ are greater than unity, leading to a relatively larger combined strength of $\gamma_{C \, (N)} = 2.06 \, (3.29)$ evaluated at the critical temperature of $T_{C \, (N)} = 119.03 \, (83.85)$ GeV. The evolution of fields for the benchmark points BP6 and BP7 given in Table~\ref{tab_bp_details} is similar in nature.


\subsection{Machine learning based classification} \label{sec_ml}

In this section, we aim to explore the possibility of distinguishing the region of parameter space that exhibits SFOPT from the region that does not, using machine learning techniques. To verify this, we employ XGBoost (eXtreme Gradient Boosting)\cite{Chen:2016btl} to classify the parameter space and eventually predict the probability of obtaining SFOPT, given a point in the considered model parameter space.

The XGBoost library equips an efficient implementation of the gradient-boosted decision tree algorithm. Decision trees can be used for both classification and regression tasks. They have a flowchart-like tree structure in which the data is recursively partitioned according to threshold conditions on input features at the internal nodes. Thus, each internal node represents a decision rule, each branch corresponds to one possible outcome of that rule, and each leaf node gives the final prediction associated with that region of feature space. In XGBoost, an additive strategy is adopted where decision trees are added one at a time, each tree improving on the previous one. To accomplish this, the algorithm defines an objective function which consists of two parts: training loss function and regularization term. The former is a measure of the model's predictability with respect to the training data while the latter involving several hyperparameters like \texttt{gamma}, \texttt{reg\_lambda} and \texttt{reg\_alpha}, controls the complexity of the model, preventing overfitting. Here we have used the \texttt{binary:logistic} objective function which is the logistic regression for binary classification. At each step, the decision tree that optimizes the objective function is added. Having briefly discussed the details of the algorithm, we now present the analysis implementing the XGBoost framework.

Among all the input parameters listed in Eq.~\ref{eq_inputs}, we train an XGBoost classifier model with the minimal set of only five input features consisting of $\{ M_2/v_r, \ M_3/v_i, \ \mu_3, \ \widetilde{\mu}_3, \ \delta_1 \}$, while ensuring that the inclusion of other features do not lead to a significant improvement in the performance of the model. The dataset used for classifier training consists of a total of 106950 samples, out of which 632 exhibit SFOPT. The three benchmark points BP1, BP4, and BP5, as given in Table~\ref{tab_bp_details}, are removed from the dataset, and the remaining is split into an 80\% training set and a 20\% test set for training and monitoring the model performance, respectively. The three benchmark points have been set aside for further validation of the probabilistic prediction of SFOPT after the training. The dataset is stratified before splitting so that the proportion of SFOPT points is almost the same in both training and testing sets. For binary classification, we define two class labels as `SFOPT' and `no-SFOPT' to identify samples based on whether the SFOPT condition is satisfied or not, respectively.

The final version of the classifier model after tuning all the hyperparameters is given below:
\begin{verbatim}
model = XGBClassifier(
    objective = `binary:logistic', n_estimators = 200, 
    learning_rate = 0.09, max_depth = 4, gamma = 1.0, max_delta_step = 1.0, 
    subsample = 0.7, scale_pos_weight = 130.0, min_child_weight = 10.0, 
    reg_alpha = 7.5, reg_lambda = 1.0
).
\end{verbatim}
\newcommand{\firstwidth}{0.395\linewidth}
\begin{figure}[t]
    \centering
    \begin{subfigure}{\firstwidth}
        \centering
        \includegraphics[width=\linewidth]{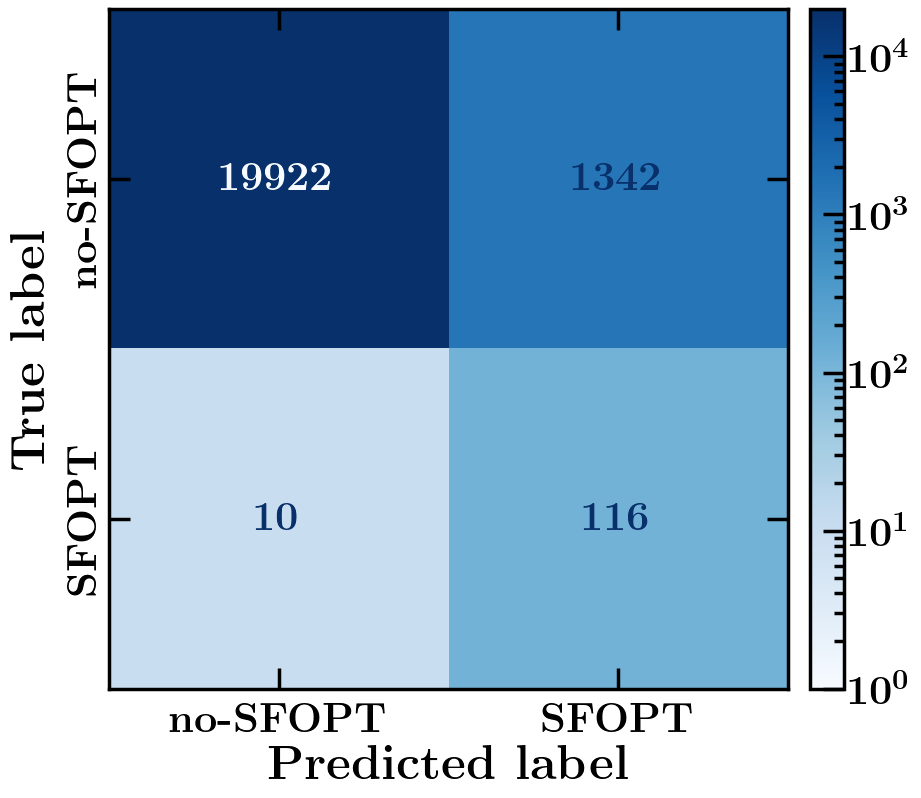}
        \caption{}
        \label{ConfusionMatrix}
    \end{subfigure} \hfill
    \begin{subfigure}{\dimexpr0.99\linewidth-\firstwidth\relax}
        \centering
        \includegraphics[width=\linewidth]{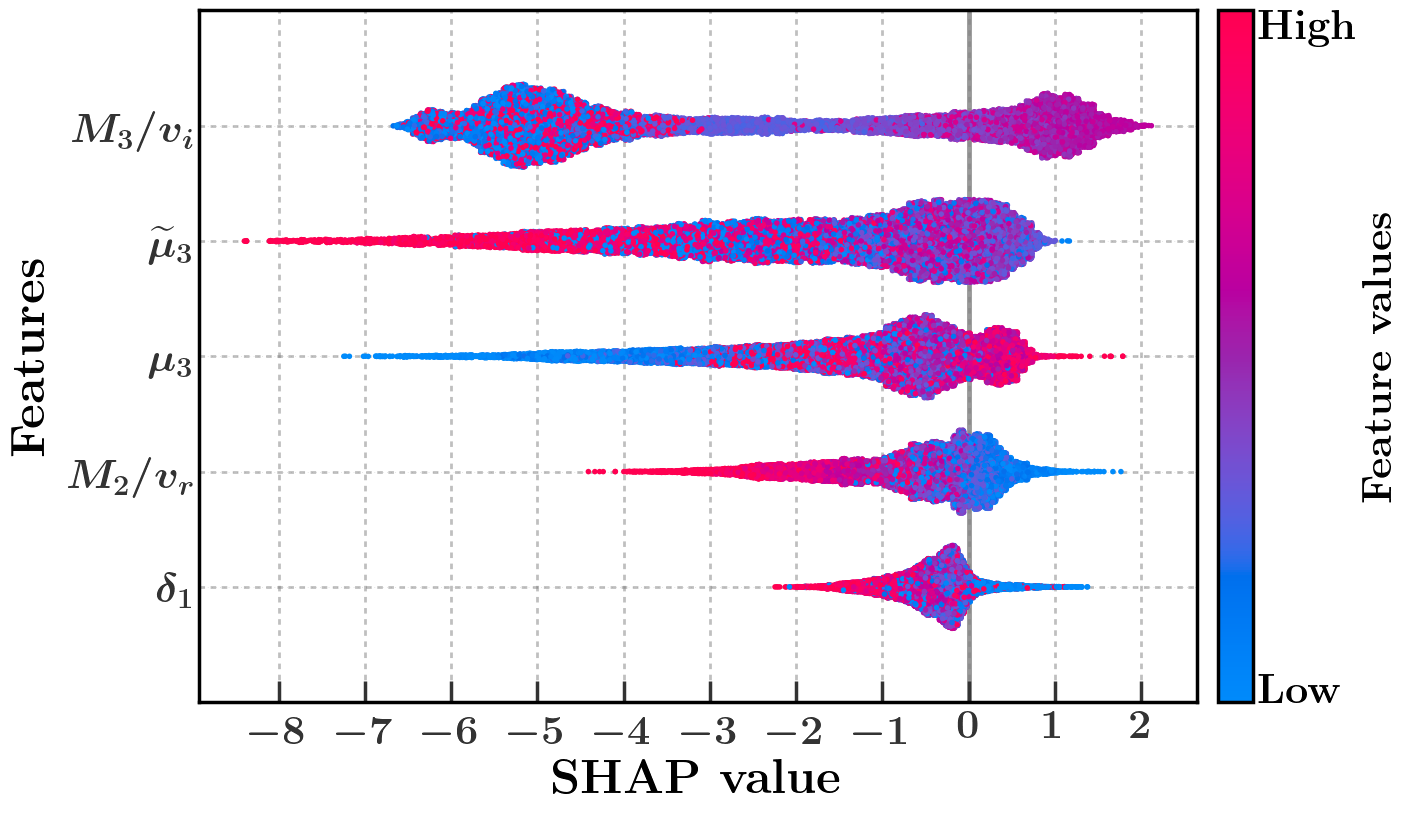}
        \caption{}
        \label{SHAPsummary}
    \end{subfigure}
    \caption{Confusion matrix (left panel) and SHAP summary plot (right panel) evaluated on the test set.}
\end{figure}
Since the dataset is highly imbalanced, with parameter points exhibiting SFOPT representing only a small fraction of the total sample, balanced accuracy (defined in Eq.~\ref{BAC}) proves to be a more meaningful performance metric than overall accuracy \cite{balacc}. Although widely used as evaluation metrics, the accuracy and area under the Receiver Operating Characteristic (ROC AUC) curve are not recommended for imbalanced datasets \cite{vodafone}. Accordingly, the hyperparameters of the classifier model are tuned to increase the balanced accuracy evaluated on the test set, also keeping an eye on the confusion matrix, thereby penalizing misclassification of SFOPT points more strongly than that of no-SFOPT points. The confusion matrix for the test set shown in Fig.~\ref{ConfusionMatrix} gives an idea of the overall performance of the model. The two diagonal elements represent correct classification of points labeled as True Negative (TN) and True Positive (TP) respectively. The off-diagonal entities are misclassified points i.e. `SFOPT' points are misclassified as `no-SFOPT' and vice versa, which are labeled as False Negative (FN) and False Positive (FP) respectively. The balanced accuracy is given as the average of recall and specificity of the model and can be directly computed from the confusion matrix given as
\begin{equation}\label{BAC}
        \text{balanced accuracy} = \frac{1}{2}\; \left(\frac{\text{TP}}{\text{TP + FN}} + \frac{\text{TN}}{\text{TN + FP}}\right) \, ,
\end{equation}
which for the above mentioned model yields $96.70\% \, (92.88 \%)$ evaluated on the training (test) set. The probabilities of exhibiting SFOPT as evaluated by the trained model for the benchmark points BP1, BP4, and BP5 given in Table~\ref{tab_bp_details}, are listed in Table~\ref{BP_Prob}, which ensures the correct classification of points yielding SFOPT to a very high degree of accuracy while remaining in consonance with the direct numerical evaluation.
\begin{table}[ht]
\centering
\begin{tabular}{lc}
\hline
\textbf{Benchmark} & \textbf{Probability of } \\
\textbf{Point} & \textbf{exhibiting SFOPT ($\%$)}\\ \hline
BP1 &  $94.23$\\
BP4 &  $90.83$\\
BP5 &  $90.87$\\ \hline
\end{tabular}
\caption{Probabilities of exhibiting SFOPT for the benchmark points BP1, BP4 and BP5 as predicted by the XGBoost classifier model.}
\label{BP_Prob}
\end{table}

To interpret the model's predictions, we use SHAP (Shapley Additive exPlanations) \cite{Lundberg:2017uca,lundberg2019consistentindividualizedfeatureattribution}, which uses a game theoretic approach to elucidate the output characteristics of generic machine learning models. The positive SHAP value of any feature tends to push the prediction towards the class `SFOPT' and vice versa, while the magnitude of the SHAP value quantifies the strength of the feature's impact on the prediction of the model. The SHAP summary plot, shown in Fig.~\ref{SHAPsummary}, is a density scatter plot to visualize the impact of all the features on the model's prediction. The features are ranked on the basis of the mean absolute SHAP value over all samples in the test dataset. Thus, the ordering of the features gives an idea of the relative average impact of the features on the model predictions.

\begin{figure}[t]
    \centering
    \begin{subfigure}{0.49\linewidth}
        \centering
        \includegraphics[width=1\textwidth]{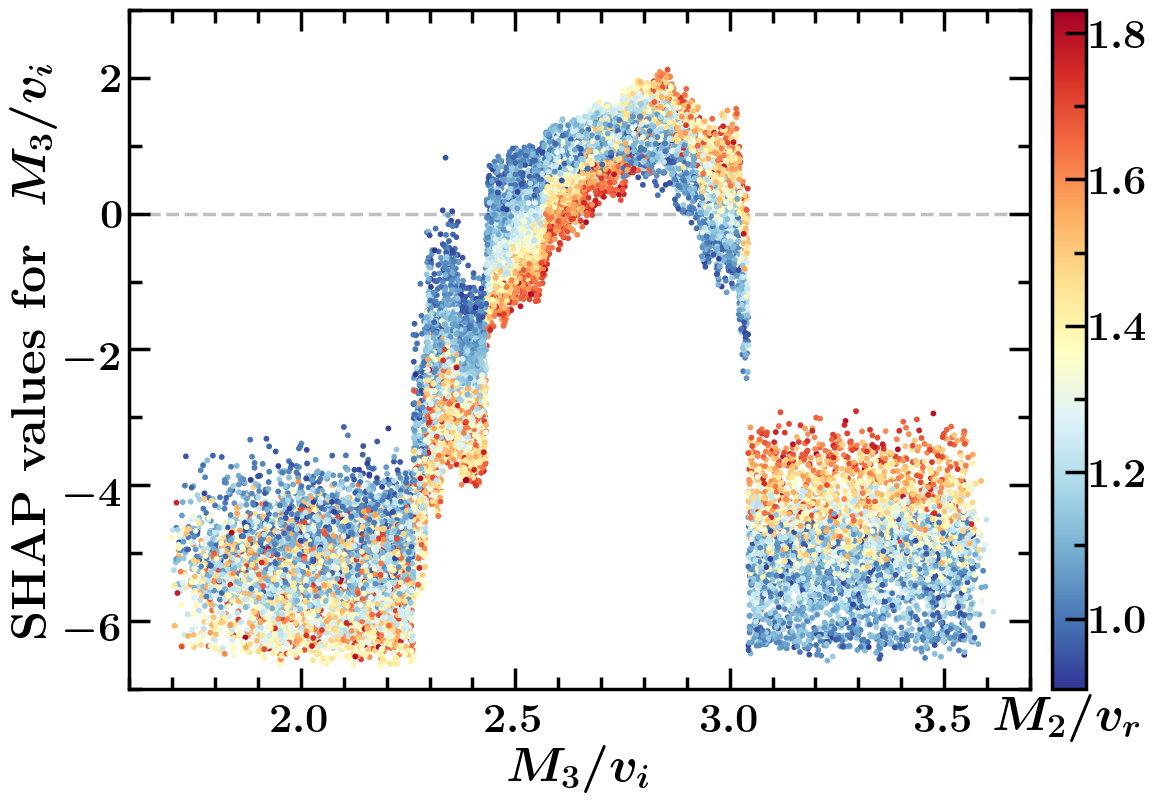}
        \caption{}
        \label{SHAPmassvev}
    \end{subfigure} \hfill
    \begin{subfigure}{0.49\linewidth}
        \centering
        \includegraphics[width=1\textwidth]{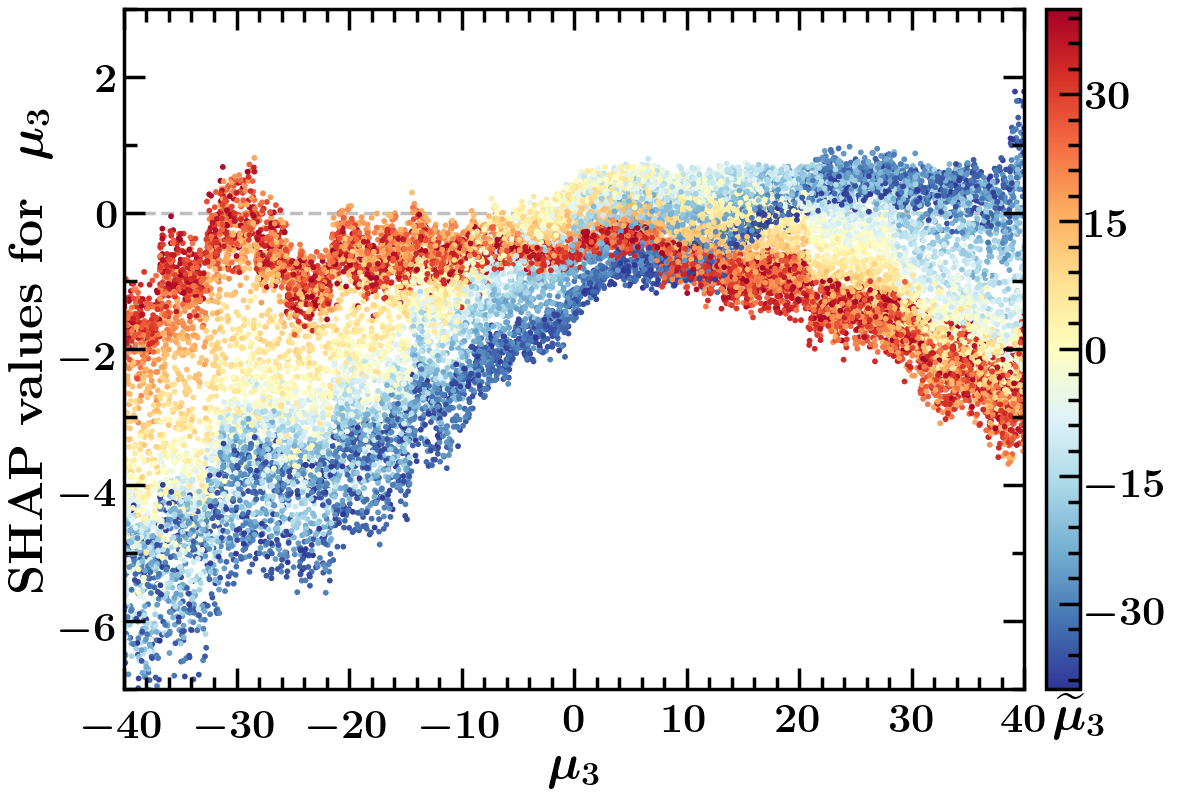}
        \caption{}
        \label{SHAPc1c2}
    \end{subfigure}
    \caption{SHAP dependence plots.}
    \label{SHAP}
\end{figure}

The dependence of the model's prediction on a specific feature can be visualized via the SHAP dependence plot, as shown in Fig.~\ref{SHAP}, while the color bar represents the values of a different feature. As can be read off from Fig.~\ref{SHAPmassvev}, the positive SHAP values lie within $2.5 \lesssim M_3/v_i \lesssim 3$, while the corresponding values of $M_2/v_r$ span the entire range, remaining consistent with Fig.~\ref{fig_param_space_mass_plane}. Additionally, the negative correlation between the two cubic terms is evident in Fig.~\ref{SHAPc1c2} where the SHAP values are predominantly positive for positive (negative) values of $\mu_3 \ (\widetilde{\mu}_3)$ in accordance with Fig~\ref{fig_param_space_c_plane}. Thus, we find that the SHAP dependence plots obtained from the classifier model lie in agreement with the numerical scans presented in section~\ref{subsec_scan_based_analysis}.

\section{Gravitational waves } \label{sec_gw}

Cosmological SFOPT in the early Universe can serve as a source of gravitational waves which carry imprints of high-energy physics operating at scales far beyond the reach of collider searches, providing a complementary probe into the thermal history of the Universe. These GWs are stochastic in nature, with three dominant mechanisms contributing to this GW background being \cite{Caprini:2015zlo, Caprini:2019egz}
\begin{enumerate}
    \item \textbf{Bubble collisions :} During an SFOPT, bubbles of the true vacuum nucleate and expand within the false vacuum. Their subsequent collisions partially convert the released vacuum energy into gravitational waves, especially in cases of runaway walls or vacuum transitions.
    
    \item \textbf{Acoustic waves :} The expanding bubble walls induce coherent motion in the surrounding plasma, generating long-lasting sound waves that efficiently source gravitational waves. This mechanism usually determines the peak frequency of the GW in first order thermal phase transitions.  

    \item \textbf{Magnetohydrodynamic (MHD) turbulence:} After the acoustic phase, the plasma motion becomes turbulent and can sustain magnetic fields, leading to an additional subdominant contribution to the stochastic GW background.  
\end{enumerate}
In general, all three mechanisms act simultaneously, and the resulting contribution to the stochastic GW background can be written as \cite{Ellis:2020awk}
\begin{equation}
\label{totgw}
    \Omega_{\rm GW} h^2 \simeq \Omega_{\rm coll} h^2 + \Omega_{\rm sw} h^2 + \Omega_{\rm turb} h^2 \, ,
\end{equation}
where the terms correspond to the contributions from bubble collisions, sound waves, and magnetohydrodynamic turbulence, respectively. The shape and amplitude of the resulting GW spectrum are governed by a few key macroscopic parameters characterizing the phase transition
\begin{itemize}
    \setlength\itemsep{0em}
    \item $\alpha$ : measures the transition strength through the vacuum energy released,
    \item $\beta/H_N$ : quantifies the inverse duration of the transition in units of the Hubble expansion rate at the nucleation epoch,
    \item $T_N$ : the nucleation temperature as defined in Eq.~\ref{nuctemp},
    \item $v_w$ : the velocity of the bubble wall that determines the efficiency of energy transfer to the plasma.
\end{itemize}
The wall velocity plays an important role in the estimation of the GW spectrum. While the precise estimation of the wall velocity \cite{Hindmarsh:2015qta, Ekstedt:2024fyq, Carena:2025flp, Si:2025vdt} for the cxSM model remains beyond the scope of this work, we choose a conservative value fixed at $v_w = 0.3$ throughout the text and present the impact of its variation on the GW spectrum in Appendix \ref{app_vw_vary_on_gw}. Among these parameters, $\alpha$ is defined as the ratio of released vacuum energy density ($\epsilon$) and total energy density of the radiation dominated Universe ($\rho_R$) evaluated at nucleation temperature and is given by \cite{Kamionkowski:1993fg}
\begin{equation}
    \alpha = \dfrac{\epsilon(T_N)}{\rho_R(T_N)} = \left. \dfrac{\Delta V_\text{eff} - T \dfrac{d \Delta V_\text{eff}}{d T}}{\dfrac{\pi^2}{30} g_*(T) T^4} \right|_{T = T_N} \, , 
\end{equation}
where $\Delta V_{\rm eff}$ corresponds to the difference between the effective potential values in the false and true vacuum with $g_{*}$ representing the number of relativistic degrees of freedom. The parameter $\beta$ quantifies the inverse time scale of the phase transition relative to the Hubble expansion rate at nucleation ($H_N$) and is given by \cite{Nicolis:2003tg}
\begin{equation}
    \frac{\beta}{H_{N}} = T \frac{d (S_{3}/T)}{dT} \bigg|_{T = T_{N}} \, , 
\end{equation}
and the contribution of bubble collisions \cite{Jinno:2016vai} in the GW spectrum within the envelope approximation, considering thin-wall bubbles, is given by
\begin{equation} \label{eq: omega_col}
\Omega_{\rm coll} h^2  = 1.67 \times 10^{-5} \, \dfrac{0.48v_w^3}{1+5.3v_w^2+5v_w^4} \, \left(\frac{100}{g_{*}(T_N)}\right)^{\frac{1}{3}}\left(\frac{H_{N}}{\beta}\right)^2 \left(\frac{\kappa_\phi \alpha}{1+\alpha}\right)^2 S_{\rm env} \left( \dfrac{f}{f_\text{env}} \right) \, ,
\end{equation}
where $\kappa_\phi$ is the efficiency factor that determines the fraction of the vacuum energy converted to the kinetic energy of the wall, $S_\text{env}$ is the spectral shape with $f_\text{env}$ being the peak frequency of the spectrum, and are given by 
\begin{align}
    \kappa_{\phi} &= \frac{1}{1 + 0.715 \alpha} \left[0.715\alpha + \frac{4}{27} \sqrt{\frac{3 \alpha}{2}}\right] \, , \\
    S_\text{env}(r) &= \left(0.064 r^{-3} + 0.456 r^{-1} + 0.48 r\right)^{-1} \, , \\
    \frac{f_\text{env}}{1\,\mu\mathrm{Hz}} &= 16.5 \left(\dfrac{0.35}{1+0.069v_w + 0.69v_w^4} \right)\left(\frac{g_{*}(T_N)}{100}\right)^{\frac{1}{6}} \left(\frac{\beta}{H_{N}}\right) \left(\frac{T_N}{100{\rm \, GeV}}\right) \, . \label{eq_peak_freq_coll}
\end{align}
Numerical simulations~\cite{Huber:2008hg, Jinno:2016vai} have shown that the efficiency of GW production from pure bubble collisions is relatively small once the plasma is included, since most of the bubble-wall energy is transferred to the surrounding fluid rather than remaining in the scalar field gradient. Consequently, $\Omega_{\rm{coll}}$ typically gives a subdominant contribution in realistic thermal phase transitions.

The contribution of sound waves in the GW spectrum redshifted to the present day Universe can be written as \cite{Hindmarsh:2017gnf}
\begin{equation} \label{eq: omega_sw}
    \Omega_{\mathrm{sw}} h^2 = 0.025 \, F_{\rm gw, 0} \, \left(\frac{\kappa_{\rm sw} \alpha}{1 + \alpha}\right)^2\, S_{\rm sw}\left( \dfrac{f}{f_\text{sw}} \right) \,  \min \left[\dfrac{H_{N}R_{N}}{\overline{U}_{f}}, 1 \right] \, \left(H_{N}R_{N}\right)h^2 \, ,
\end{equation}
where $F_{\rm gw,0} = 3.57 \times 10^{-5} \left({100}/{g_{*}(T_N)}\right)^{1/3}$, $\kappa_{sw}$ is the efficiency factor, $\overline{U}_{f}$ is the enthalpy-weighted root mean square of the fluid velocity and the mean bubble radius is given by $R_{N} = \left(8 \pi\right)^{1/3} \left (v_w / \beta \right)$ at the nucleation temperature \cite{Athron:2024xrh}. The spectral shape and peak frequency are respectively given by 
\begin{align}
    S_{\rm sw}(r) &= r^3 \left(\frac{7}{4 + 3 r^2}\right)^{7/2} \, \\
    \frac{f_{\rm sw}}{1\,\mu\mathrm{Hz}} &= 2.6  \left(\frac{T_{N}}{100\mathrm{\, GeV}}\right) \left(\frac{g_{*}(T_N)}{100}\right)^{\frac{1}{6}} \left(\frac{1}{H_{N}R_{N}}\right) \, . \label{eq_peak_freq_sw}
\end{align}
The dominant source of gravitational waves usually arises from long-lasting acoustic (sound) waves in the plasma. During bubble expansion and collision, the kinetic energy of the walls is efficiently transferred to the surrounding fluid, producing bulk motion and coherent sound waves that persist for about a Hubble time after the transition. Since these sound waves last much longer than the brief collision phase, they typically dominate the total gravitational wave energy density in most electroweak-scale first-order phase transitions.

The part of the GW spectrum resulting from the MHD turbulence, redshifted to today, is given by \cite{Binetruy:2012ze}
\begin{equation} \label{eq: omega_turb}
\Omega_{\mathrm{turb}} h^2={} 3.35 \times 10^{-4} \, v_w \, \left(\frac{H_N}{\beta}\right)\left(\frac{\kappa_{\mathrm{turb}} \alpha}{1+\alpha}\right)^{3 / 2} \left(\frac{100}{g_*(T_N)}\right)^{1 / 3}
\frac{{(f / f_{\mathrm{turb}}})^3}{(1 + f / f_{\mathrm{turb}})^{11 / 3}\left(1+8 \pi f / H_0\right)},
\end{equation}
where $\kappa_{turb}$ is chosen as a small fraction of $\kappa_{sw}$ \cite{Hindmarsh:2015qta}. The peak frequency and the red-shifter Hubble rate at the production of GWs are, respectively, given by 
\begin{align}
    \dfrac{f_{\mathrm{turb}}}{1 \, \mu\mathrm{Hz}} &= 27 \frac{1}{v_w}\left(\frac{\beta}{H_N}\right)\left(\frac{T_{N}}{100{\rm GeV}}\right)\left(\frac{g_*(T_N)}{100}\right)^{1 / 6} \, , \label{eq_peak_freq_turb} \\
    H_0 &= 16.5 \, \left(\frac{g_{*} (T_N)}{100}\right)^{\frac{1}{6}} \left(\frac{T_{N}}{100 {\rm \, GeV}}\right)~\mu\mathrm{Hz} \, .
\end{align}
The contribution arising from magnetohydrodynamic (MHD) turbulence is generated in the plasma once the acoustic waves become non-linear and turbulent. A small fraction of the fluid kinetic energy cascades into turbulent vortices and possibly magnetic fields, which can further source stochastic GWs \cite{Caprini:2009yp}. This component is typically suppressed compared to the sound-wave term, since only a minor fraction of the bulk kinetic energy is converted into turbulence. Nevertheless, it can enhance the high-frequency tail of the spectrum and is relevant for probing small-scale plasma dynamics.

\begin{figure}[t]
    \centering
    \begin{subfigure}{0.3275\linewidth}
        \centering
        \includegraphics[width=1\textwidth]{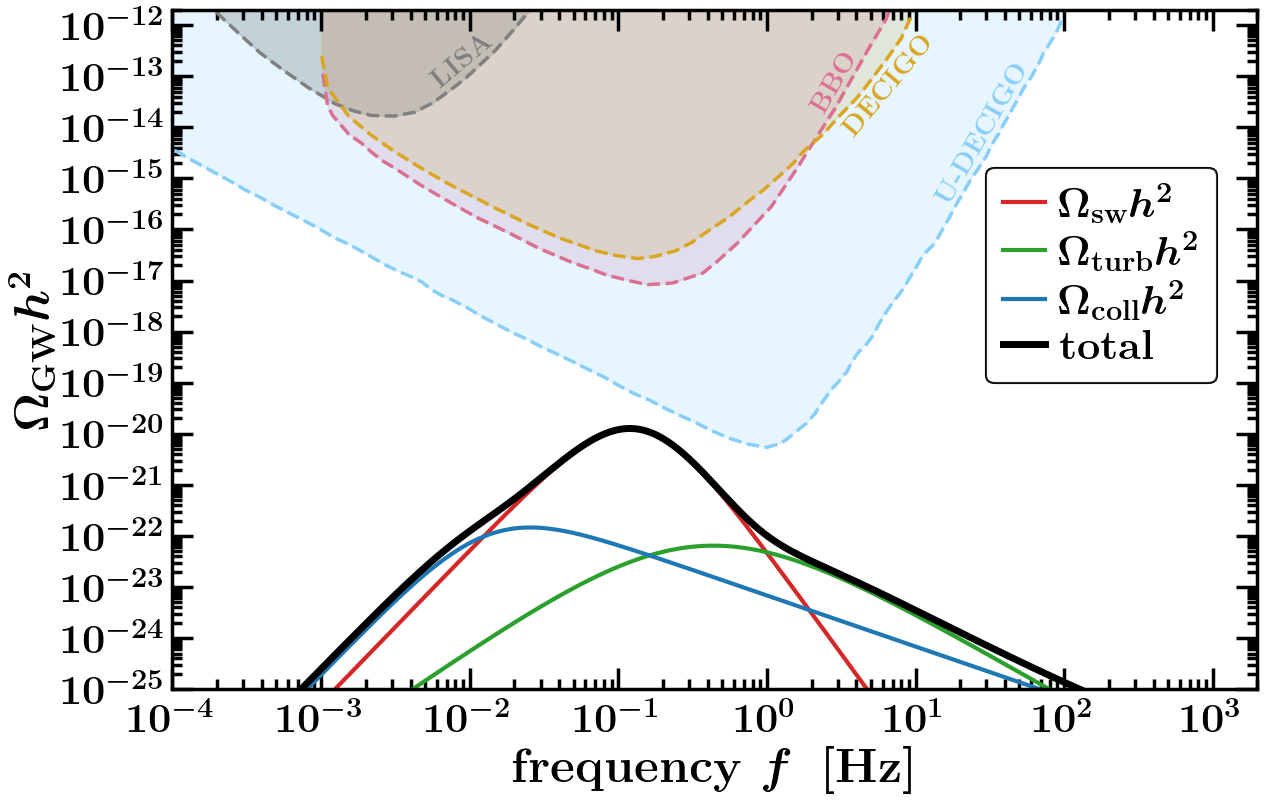}
        \caption{}
        \label{fig_bp1_gw}
    \end{subfigure} \hfill
    \begin{subfigure}{0.3275\linewidth}
        \centering
        \includegraphics[width=1\textwidth]{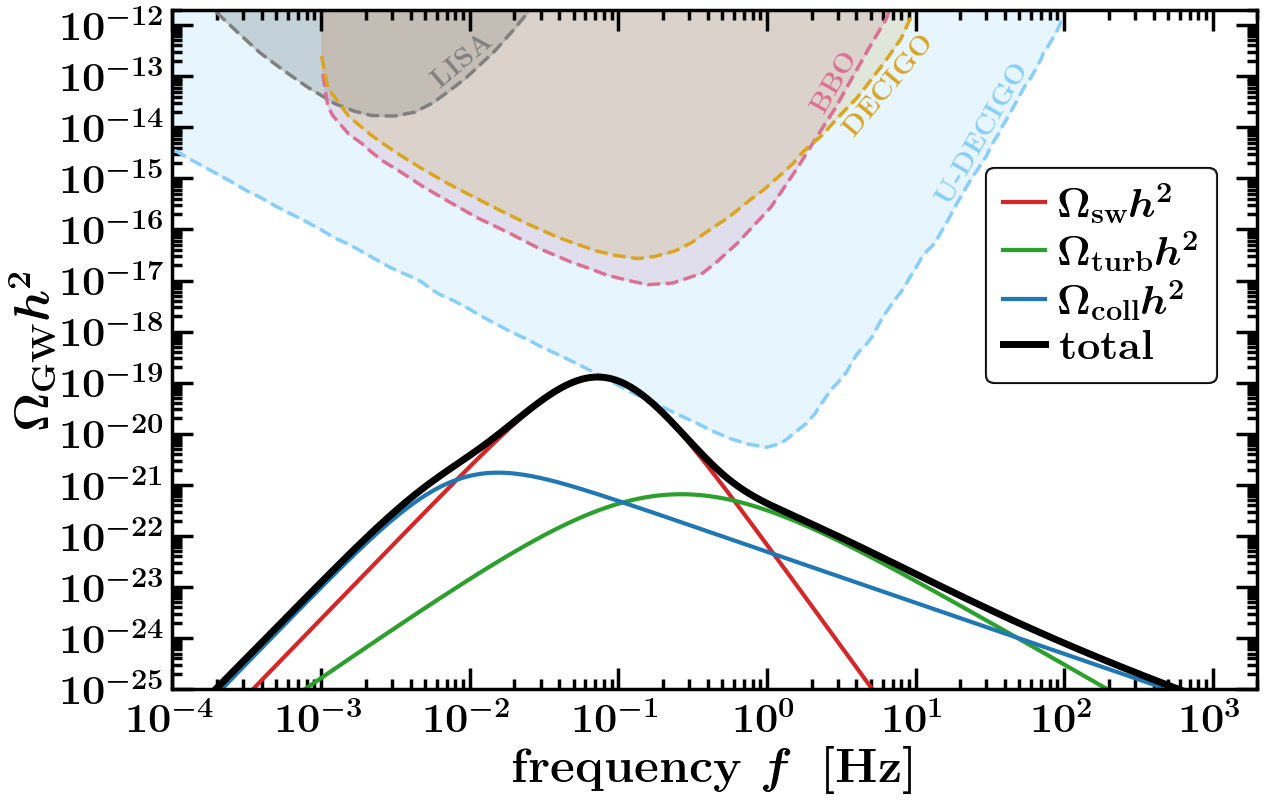}
        \caption{}
        \label{fig_bp2_gw}
    \end{subfigure} \hfill
    \begin{subfigure}{0.3275\linewidth}
        \centering
        \includegraphics[width=1\textwidth]{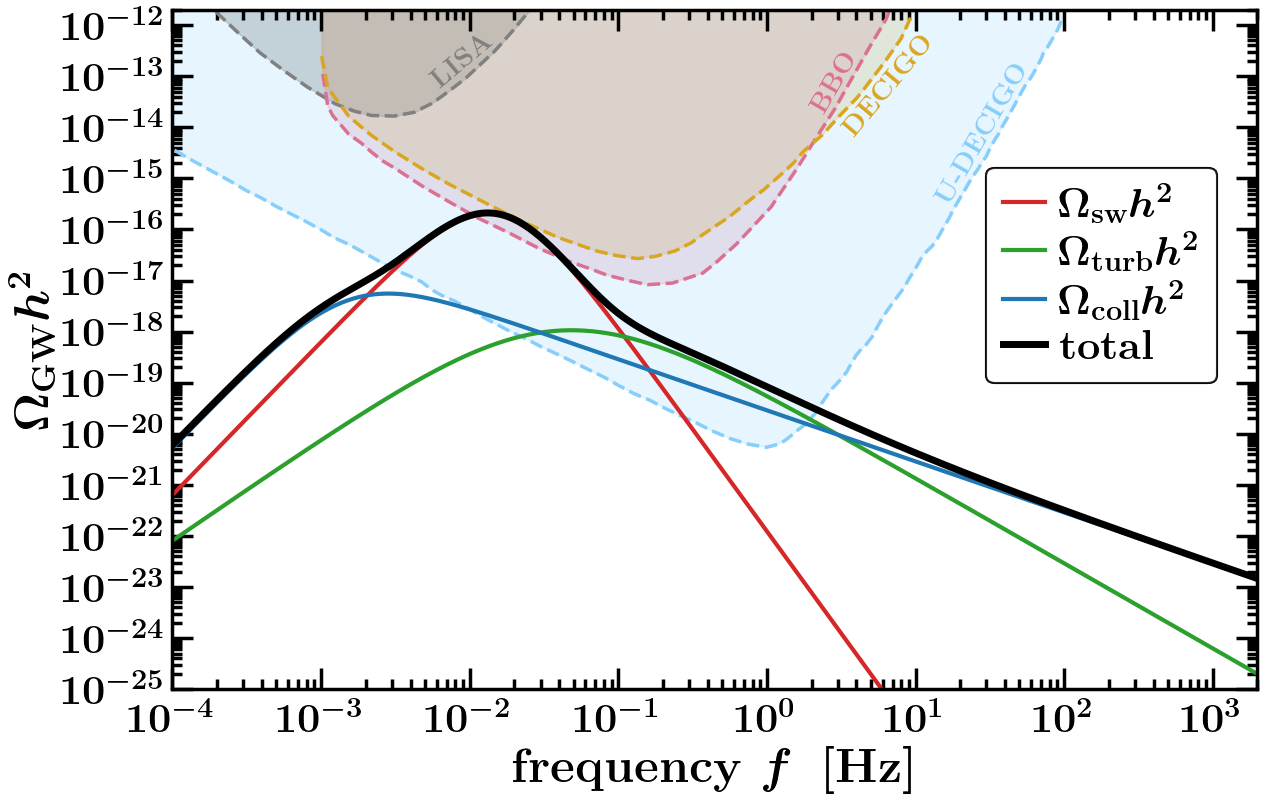}
        \caption{}
        \label{fig_bp3_gw}
    \end{subfigure}
    \caption{The stochastic GW spectrum for BP1 (left panel), BP4 (middle panel) and BP5 (right panel). Individual contributions for each benchmark point from sound wave, magnetohydrodynamic turbulence, and bubble collisions are shown in red, green, and blue solid lines corresponding to Eq.~\ref{eq: omega_sw}, \ref{eq: omega_turb}, and \ref{eq: omega_col}, respectively. The dashed gray, violet, yellow, and light blue lines correspond to the sensitivity curves for LISA, BBO, DECIGO, and U-DECIGO, respectively.}
\end{figure}

In Figs.~\ref{fig_bp1_gw}, \ref{fig_bp2_gw} and \ref{fig_bp3_gw}, we present the GW spectra corresponding to strong FOPT occurring along the BSM direction (BP1), along the SM direction (BP4), and along both directions (BP5), respectively. For each benchmark point, we show the individual contributions of each term in Eq.~\ref{totgw}. As expected, the acoustic contribution is the highest, the collision contribution is higher in the lower frequency side, while that of MHD is higher in the higher frequencies. The physical reason is that, during a first-order phase transition, the expanding bubble walls interact strongly with the surrounding relativistic plasma, transferring most of their energy into bulk fluid motion rather than remaining as scalar-field gradients \cite{Caprini:2019egz}. Unlike the long-lived acoustic source, gravitational waves from bubble collisions are produced only for a brief moment during the overlap of expanding walls. The turbulence in the plasma that follows contributes at higher frequencies but carries a comparatively small and rapidly fading share of the released energy.
\begin{figure}
    \centering
    \includegraphics[width=0.5\linewidth]{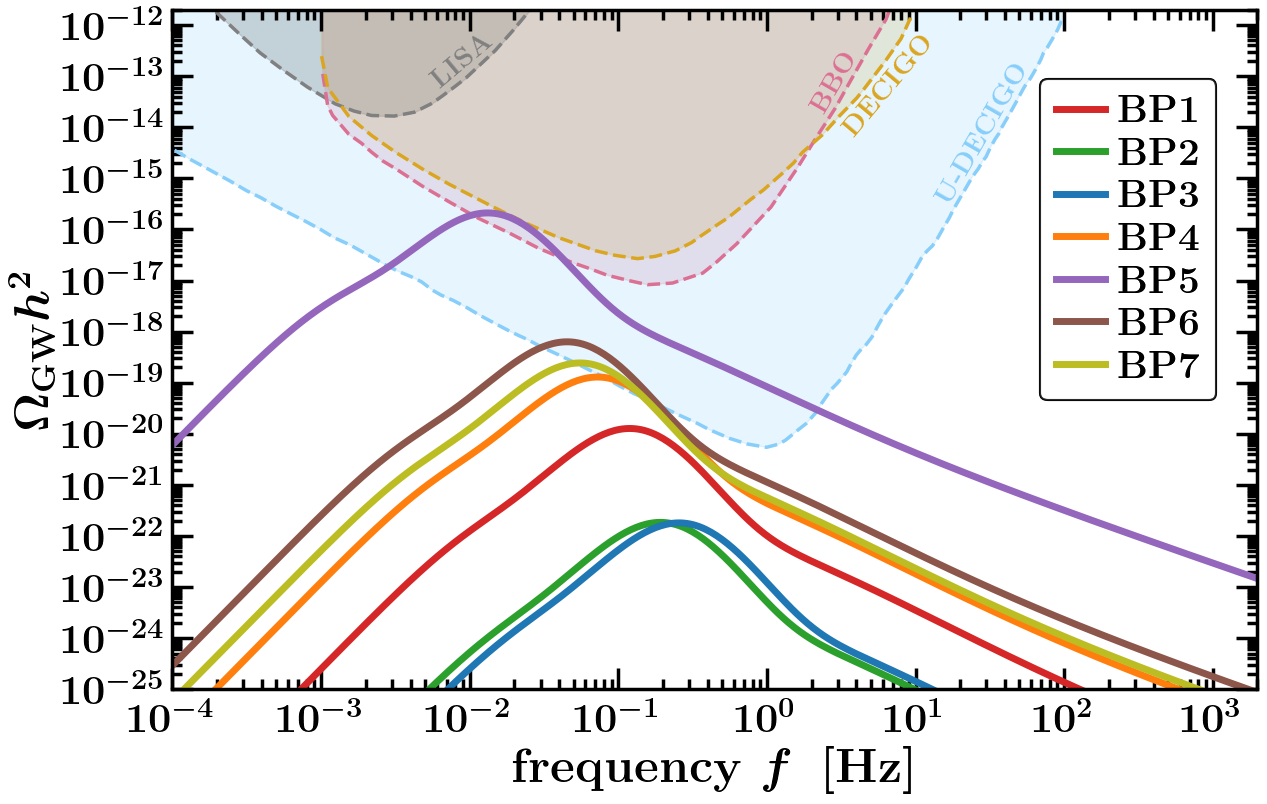}
    \caption{GW spectrum for all the benchmark points with the sensitivity curves for different experiments.}
   \label{fig_all_gw}
\end{figure}

Finally, we show the GW spectrum for all the benchmark points in Fig. \ref{fig_all_gw}. The peak of the GW spectrum is highest for BP5, followed by BP6 and then BP7, where SFOPT is observed along both directions. The GW spectrum due to BP4 corresponds to the case where SFOPT is observed only along the SM direction with a lower peak than that of BP5, BP6, and BP7, as expected.  Similarly, for the spectrum corresponding to the BSM directions, the peak is highest for BP1, followed by BP3, and then BP2, and they are even lower than that of the SM case. This is due to the combined effect of the strength of the phase transition $\alpha$ and the inverse duration of the phase transition ${\beta}/{H_{N}}$ as can be verified from Table \ref{tab_bp_output}. Almost all GW spectra corresponding to the benchmarks where SFOPT is observed along both directions can be detected in the upcoming BBO, DECIGO, and U-DECIGO  detectors.


\section{Conclusion and discussion} \label{sec_conclusion}

In this paper, we have revisited one of the simplest extensions to the Standard Model by incorporating a complex singlet scalar, without imposing any extra discrete or continuous global symmetries on this scalar singlet field. This approach allows for a fully general treatment of the scalar potential, enabling richer possibilities in the vacuum structure. Specifically, it allows a configuration where both the real and imaginary components of the singlet acquire vevs alongside the Standard Model Higgs field, leading to a more intricate interplay among the scalars.

We first imposed various theoretical and experimental constraints on the parameters of the full scalar potential. To explore the thermal history of the early Universe, we have calculated the finite temperature effective potential at the one-loop level, incorporating daisy resummation and counter terms. This framework enabled us to investigate the electroweak phase transition dynamics across all three relevant field directions i.e. the SM Higgs and the two singlet components. Adopting a hierarchical mass spectrum for the physical scalars to align with phenomenological viability, we have conducted a comprehensive parameter scan over the model's 11 free parameters. This effort identified viable regions of the parameter space capable of supporting a strong first-order phase transition (SFOPT), a key ingredient for successful electroweak baryogenesis and the generation of observable gravitational waves. We have classified the phase transitions based on the primary field directions involved in the SFOPT, resulting in three distinct categories: those dominated by the Standard Model direction, those along the beyond-Standard-Model singlet directions, and hybrid cases involving both. To illustrate these behaviors, we have chosen a few representative benchmark points, some of which exhibit multi-step transitions characterized by sequential symmetry breaking at different critical temperatures, with vev-to-temperature ratios exceeding unity at the critical temperature, confirming the strength of the transitions. Through our scan of the model parameter space, we have shown some insightful patterns: SFOPT tends to occur in regimes where the ratios of the masses of the heavier scalars to their corresponding singlet vevs are close to order one. More precisely, the parameter space favors $M_2/v_r $ falling between approximately 1.0 and 1.5 for the real singlet component, and $M_3/v_i$ approximately between 2.5 and 3.0 for the imaginary part, highlighting the delicate balance required between mass scales and couplings. We have also explored how well a machine learning model can predict the probability of yielding SFOPT in regions of parameter space via the XGBoost classifier. We have also verified that the trained model is capable of predicting SFOPT at representative benchmark points with probabilities exceeding 90\%.

Furthermore, we have computed the stochastic gravitational wave spectra arising from bubble nucleation and related dynamics in these benchmark scenarios. The spectra are shaped by contributions from sound waves in the plasma, magnetohydrodynamic turbulence, and bubble wall collisions, with the sound wave component often dominating because of efficient energy transfer in relativistic fluids. Particularly, the gravitational wave signals from points featuring SFOPT along both Standard Model and beyond-Standard-Model directions exhibit peak amplitudes and frequencies that intersect the projected sensitivity curves of advanced detectors like DECIGO, as well as potentially BBO and U-DECIGO. This overlap suggests that such signals could be detectable in the near future, providing a complementary probe to collider experiments and opening a new window into high-energy physics beyond the Standard Model.

Overall, the SM augmented with a complex singlet scalar scenario appears to be a versatile and minimal framework that not only addresses shortcomings in the Standard Model, such as the baryon asymmetry, but also predicts testable multimessenger signatures. Future work could extend this analysis to incorporate CP violation mechanisms for full baryogenesis calculations or explore collider signatures of the additional scalars to further constrain the viable parameter space.


\section*{Acknowledgments}

RP would like to thank Peter Athron, Andrew Fowlie, and Yang Zhang for helping with the use of \texttt{PhaseTracer2}. RP and DM acknowledges Sougata Ganguly and Amit Adhikary for various discussions. DM and KM acknowledge the financial support provided by the Indian Association for the Cultivation of Science (IACS), Kolkata.


\begin{appendix}
\renewcommand{\thesection}{\Alph{section}}
\renewcommand{\theequation}{\thesection-\arabic{equation}}
\setcounter{equation}{0}


\section{Lagrangian parameters in terms of physical quantities} \label{app_lag_to_physical}
Here we show the functional dependence of quartic couplings mentioned in the tree-level potential given in Eq. \ref{eq_Vtree} in terms of physical mass eigenvalues, mixing angles, and vevs. Utilizing the tadpole equations given in \ref{eq_tadpole1}, \ref{eq_tadpole2} and \ref{eq_tadpole3} the quartic couplings can be expressed as
\begin{align}
    \lambda &= \dfrac{2}{v_h^2} A \, , \\
    \delta_2 &= \dfrac{1}{4 v_h v_r v_i} \left(4 B v_i + 4 C v_r - \roottwo v_h v_i \delta_1 \right) \, , \\
    \delta_3 &= \dfrac{1}{4 v_h v_r v_i} \left(4 B v_i - 4 C v_r - \roottwo v_h v_i \delta_1 \right) \, , \\
    d_1 &= \dfrac{1}{ 96 v_i^2 v_r^3} \bigg[ 24 \roottwo a_1 v_i^2 + 2 \roottwo \widetilde{\mu}_3 v_i^2 (v_r^2 + v_i^2) - 6 \roottwo \mu_3 v_i^2 (v_i^2 + 5 v_r^2) \, \nonumber \\
        &+ 3 (8 D v_i^2 v_r - 16 E v_i v_r^2 + 8 F v_r^3 + \roottwo v_h^2 v_i^2 \delta_1) \bigg] \, , \\
    d_2 &= \dfrac{1}{96} \bigg[ \dfrac{72 F}{v_i^2} + \dfrac{72 D}{v_r^2} + \dfrac{48 E + 2 \roottwo (3 \mu_3 - 13 \widetilde{\mu}_3) v_i}{v_i v_r} + \dfrac{3 \roottwo }{v_r^3} \left(24 a + 2 (-3 \mu_3 + \widetilde{\mu}_3) v_i^2 + 3 v_h^2 \delta_1 \right) \bigg] \, , \\
    d_3 &= - \dfrac{F}{v_r^2} + \dfrac{D}{v_r^2} - \dfrac{\mu_3 + \widetilde{\mu}_3}{2 \roottwo v_r} + \dfrac{1}{12 \roottwo v_r^3} \left( 24 a + 2 (-3 \mu_3 + \widetilde{\mu}_3) v_i^2 + 3 v_h^2 \delta_1 \right) \, ,
\end{align}
where
\begin{eqnarray}
\begin{aligned}
A &= M_1^2 c_2^2 c_3^2 + M_2^2 (c_3 s_1 s_2 + c_1 s_3)^2 + M_3^2 (c_1 c_3 s_2 - s_1 s_3)^2 \, , \\ 
B &= (M_3^2 - M_2^2) \cos(2 \theta_3) c_1 s_1 s_2 + c_1^2 c_3 (M_3^2 s_2^2 - M_2^2) s_3 + c_3 (M_1^2 c_2^2 + s_1^2 (M_2^2 s_2^2 - M_3^2)) s_3 \, , \\
C &= c_2 (c_3 (M_1^2 - M_3^2 c_1^2 - M_2^2 s_1^2) s_2 + (-M_2^2 + M_3^2) c_1 s_1 s_3) \, , \\
D &= M_1^2 c_2^2 s_3^2 + M_3^2 (c_3 s_1 + c_1 s_2 s_3)^2 + M_2^2 (c_1 c_3 - s_1 s_2 s_3)^2 \, , \\
E &= c_2 ((M_2^2 - M_3^2) c_1 c_3 s_1 + (M_1^2 - M_3^2 c_1^2 - M_2^2 s_1^2) s_2 s_3) \, , \\
F &= c_2^2 (M_3^2 c_1^2 + M_2^2 s_1^2) + M_1^2 s_2^2 \, .
\end{aligned}
\end{eqnarray}


\section{Counter terms in the effective potential} \label{app_counter_terms}

Utilizing the renormalization condition mentioned in Eq. \ref{eq_renorm_cond}, we fix the coefficients of the counter terms given in Eq. \ref{eq_Vct} and are given by
\newcommand{\dv}[1]{V_\text{CW}^{#1}}
\begin{align}
    \delta \mu_H^2 &= \dfrac{1}{v_h^2 v_i^3} \bigg[  -v_h v_i^3 \left(-3 \dv{h} + v_h \dv{h, h} + v_i \dv{h, s_i} \right) - v_i^3 v_r \left(2 \dv{s_r} + v_h \dv{h, s_r} - 2 v_r \dv{s_r, s_r} \right) \nonumber \\
    &+ 2 v_r^4 \left(\dv{s_i} - v_i \dv{s_i, s_i} \right) \bigg] \, , \\
    \delta \lambda &= \dfrac{2}{v_h^3} \bigg[ \dv{h} - v_h \dv{h, h} \bigg] \, , \\
    \delta \delta_1 &= \dfrac{4 \roottwo}{v_h^2 v_i^3} \bigg[ - v_i^3 \dv{s_r} + v_i^3 v_r \dv{s_r, s_r} + v_r^3 \left(\dv{s_i} - v_i \dv{s_i, s_i} \right) \bigg] \\
    \delta \delta_2 &= \dfrac{1}{v_h^2 v_i^3 v_r} \bigg[ v_i^3 \left( 2 \dv{s_r} - v_h \dv{h, s_r} \right) - v_i^2 v_r \left( v_h \dv{h, s_i} + 2 v_i \dv{s_r, s_r} \right)  - 2 v_r^3 \left(\dv{s_i} - v_i \dv{s_i, s_i} \right)\bigg] \, , \\
    \delta \delta_3 &= \dfrac{1}{v_h^2 v_i^3 v_r} \bigg[ v_i^3 \left( 2 \dv{s_r} - v_h \dv{h, s_r} \right) + v_i^2 v_r \left( v_h \dv{h, s_i} - 2 v_i \dv{s_r, s_r} \right)  - 2 v_r^3 \left(\dv{s_i} - v_i \dv{s_i, s_i} \right)\bigg] \, , \\
    \delta b_1 &= \dfrac{1}{2 v_i^3 v_r} \bigg[ vi^3 \left(-2 \dv{s_r} + v_h \dv{h, s_r} + v_i \dv{s_r, s_i} \right) - vi^2 v_r \left(-3 \dv{s_i} + v_h \dv{h, s_i} + v_i \dv{s_i, s_i} \right) \nonumber \\
    &- v_i^2 v_r^2 \dv{s_r, s_i} + v_r^3 \left(- \dv{s_i} + v_i \dv{s_i, s_i} \right) \bigg] \, , \\
    \delta b_2 &= \dfrac{1}{2 v_i^3 v_r} \bigg[ vi^3 \left(-2 \dv{s_r} + v_h \dv{h, s_r} + v_i \dv{s_r, s_i} \right) + vi^2 v_r \left(-3 \dv{s_i} + v_h \dv{h, s_i} + v_i \dv{s_i, s_i} \right) \nonumber \\
    &+ v_i^2 v_r^2 \dv{s_r, s_i} + v_r^3 \left(- \dv{s_i} + v_i \dv{s_i, s_i} \right) \bigg] \, , \\
    \delta d_1 &= \dfrac{1}{2 v_i^3 v_r} \bigg[ v_i^2 \dv{s_r, s_i} + v_r \dv{s_i} - v_i v_r \dv{s_i, s_i} \bigg] \, , \\
    \delta d_2 &= \dfrac{1}{2 v_i^3 v_r} \bigg[ - v_i^2 \dv{s_r, s_i} + 3 v_r \dv{s_i} - 3 v_i v_r \dv{s_i, s_i} \bigg] \, ,
\end{align}
with the shorthand notation of the partial derivatives given as
\begin{equation}
    \dv{x} \equiv \dfrac{\partial V_\text{CW}}{\partial x} \, ; \quad \dv{x, y} \equiv \dfrac{\partial^2 V_\text{CW}}{\partial x \partial y} \, ,
\end{equation}
has been used where $x, y \in \lbrace h, s_r, s_i \rbrace$. Without an adequate number of renormalization conditions, all other counter terms appearing in Eq.~\ref{eq_Vct} have been chosen to be equal to zero.


\section{Renormalization scale dependency of GW parameters} \label{app_mu_vary_on_gw}

\begin{figure}[h]
    \centering
    \begin{subfigure}{0.32\textwidth}
        \centering
        \includegraphics[width=1\linewidth]{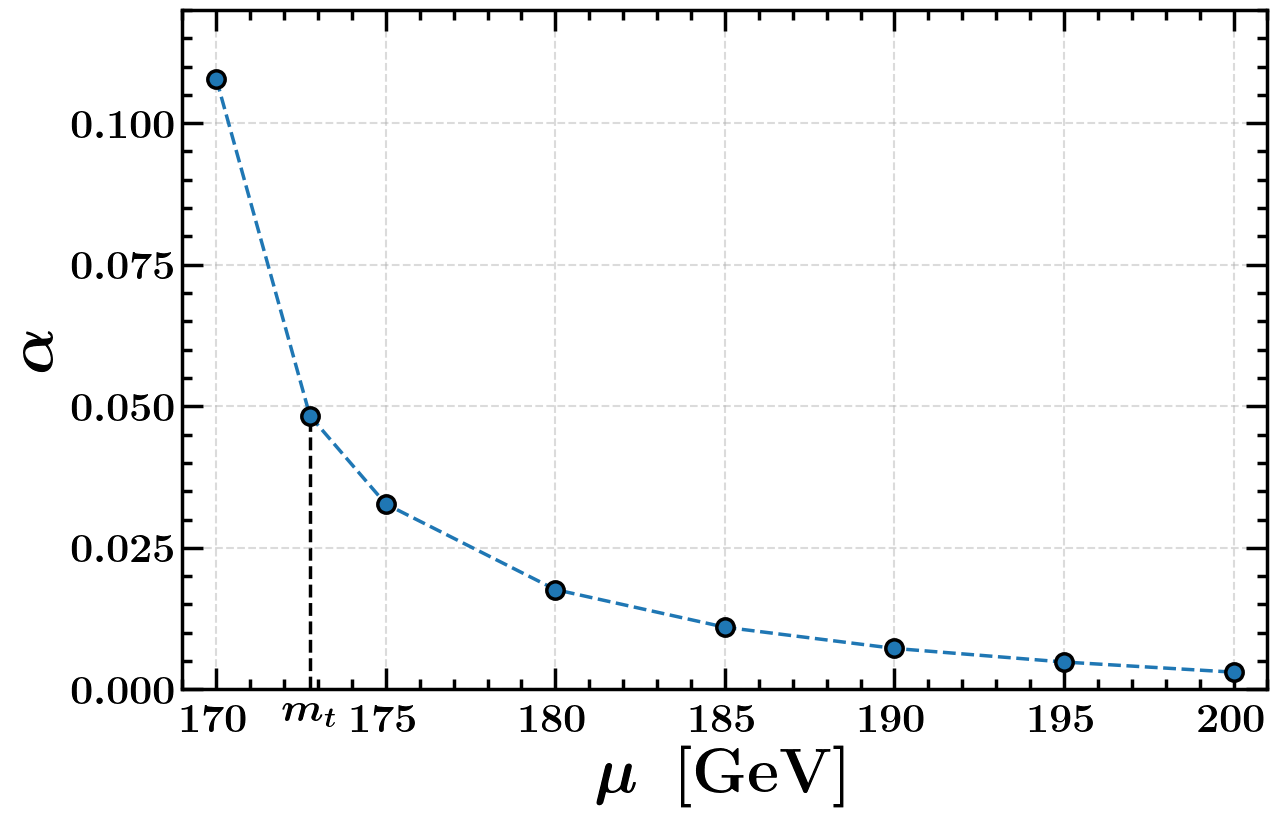}
        \caption{}
        \label{fig_mu_vary_alpha}
    \end{subfigure}
    \begin{subfigure}{0.32\textwidth}
        \centering
        \includegraphics[width=1\linewidth]{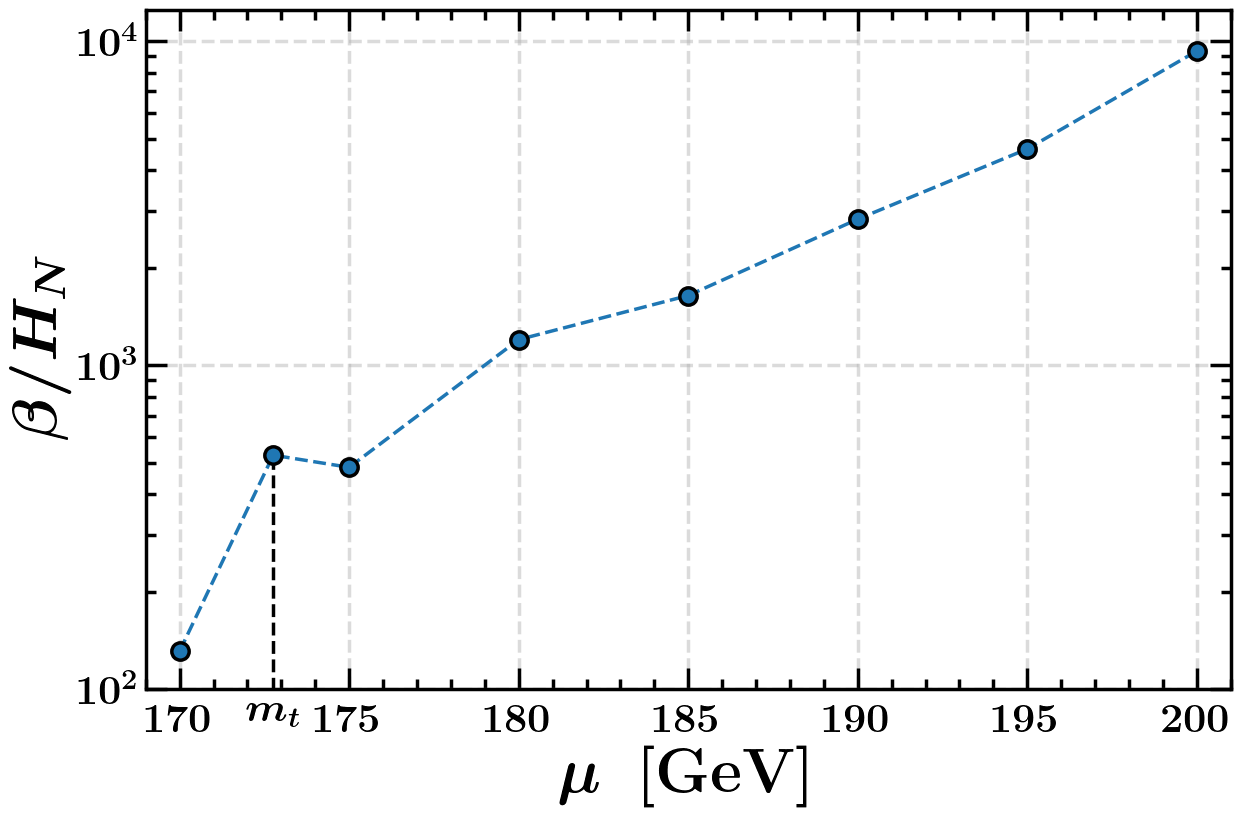}
        \caption{}
        \label{fig_mu_vary_beta}
    \end{subfigure}
    \begin{subfigure}{0.32\textwidth}
        \centering
        \includegraphics[width=1\linewidth]{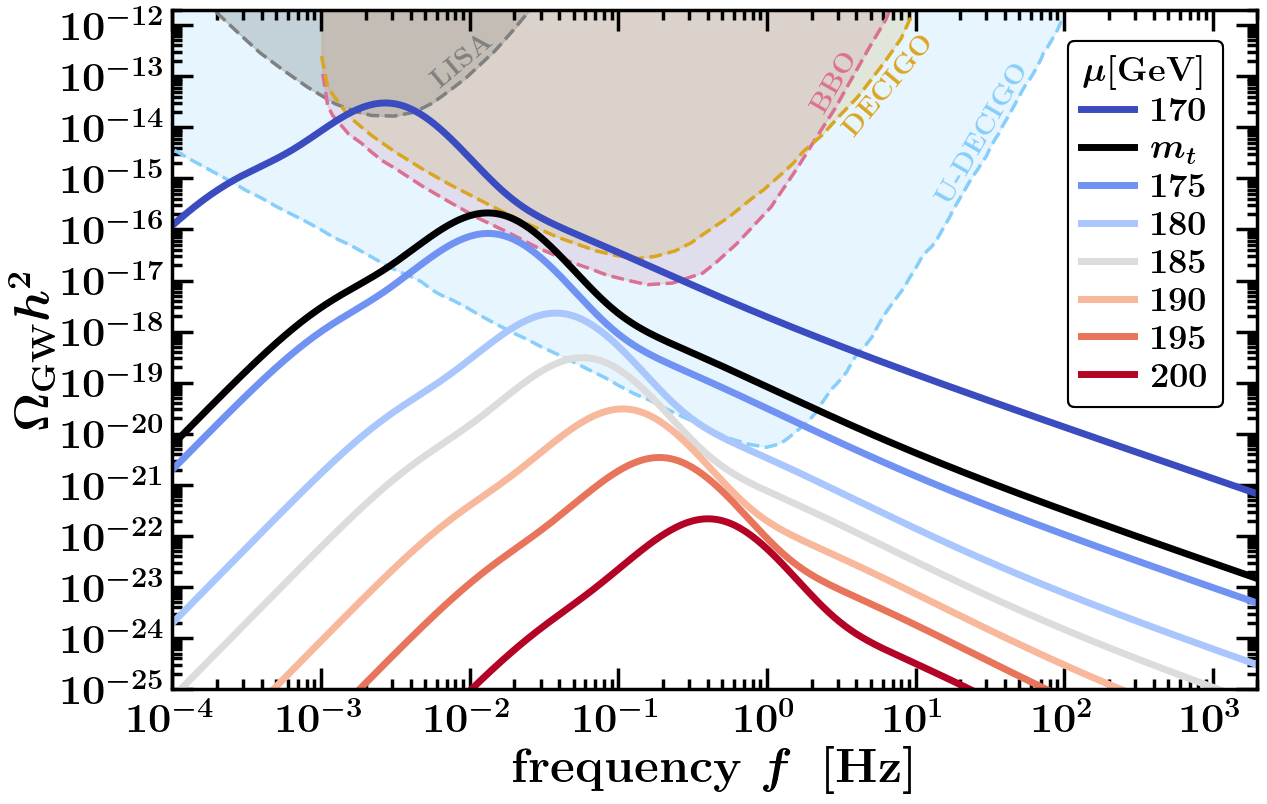}
        \caption{}
        \label{fig_mu_vary_gw}
    \end{subfigure}
    \caption{Variation of $\alpha$, $\beta$ and GW spectrum with respect to the renormalization scale $\mu$ as shown in the left (\ref{fig_mu_vary_alpha}), middle (\ref{fig_mu_vary_beta}) and right (\ref{fig_mu_vary_gw}) panels respectively for the benchmark point BP5.}
    \label{fig_mu_vary}
\end{figure}

\noindent
It has been well explored that the GW parameters extracted from one loop effective potential exhibits significant dependence on the renormalization scale \cite{Croon:2020cgk, Gould:2021oba}. The dependence can be made milder by going beyond one-loop calculations. In this appendix, we choose a benchmark point with the highest strength of SFOPT, i.e BP5, and show the dependence of $\alpha, \beta$ and the resulting GW spectrum on $\mu$ in Fig. \ref{fig_mu_vary}. Without the inclusion of higher order effects in the couplings, we notice an decreasing and increasing nature of $\alpha$ and $\beta$ respectively with respect to $\mu$, which is in agreement with the findings obtained earlier \cite{Croon:2020cgk, Gould:2021oba}.


\section{Impact of the wall velocity on GW spectrum} \label{app_vw_vary_on_gw}

\begin{figure}[h]
    \centering
    \begin{subfigure}{0.32\textwidth}
        \centering
        \includegraphics[width=1\linewidth]{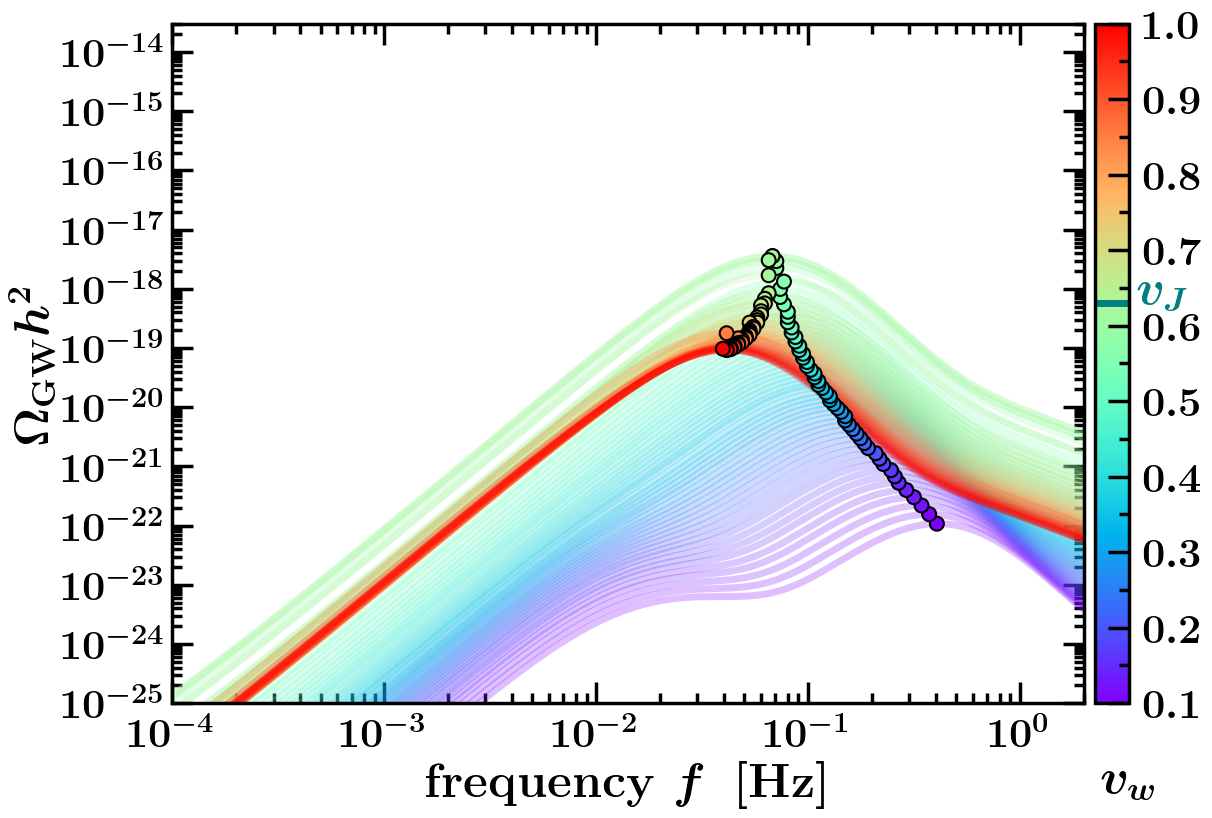}
        \caption{}
        \label{fig_vw_vary_bp1_bsm}
    \end{subfigure}
    \begin{subfigure}{0.32\textwidth}
        \centering
        \includegraphics[width=1\linewidth]{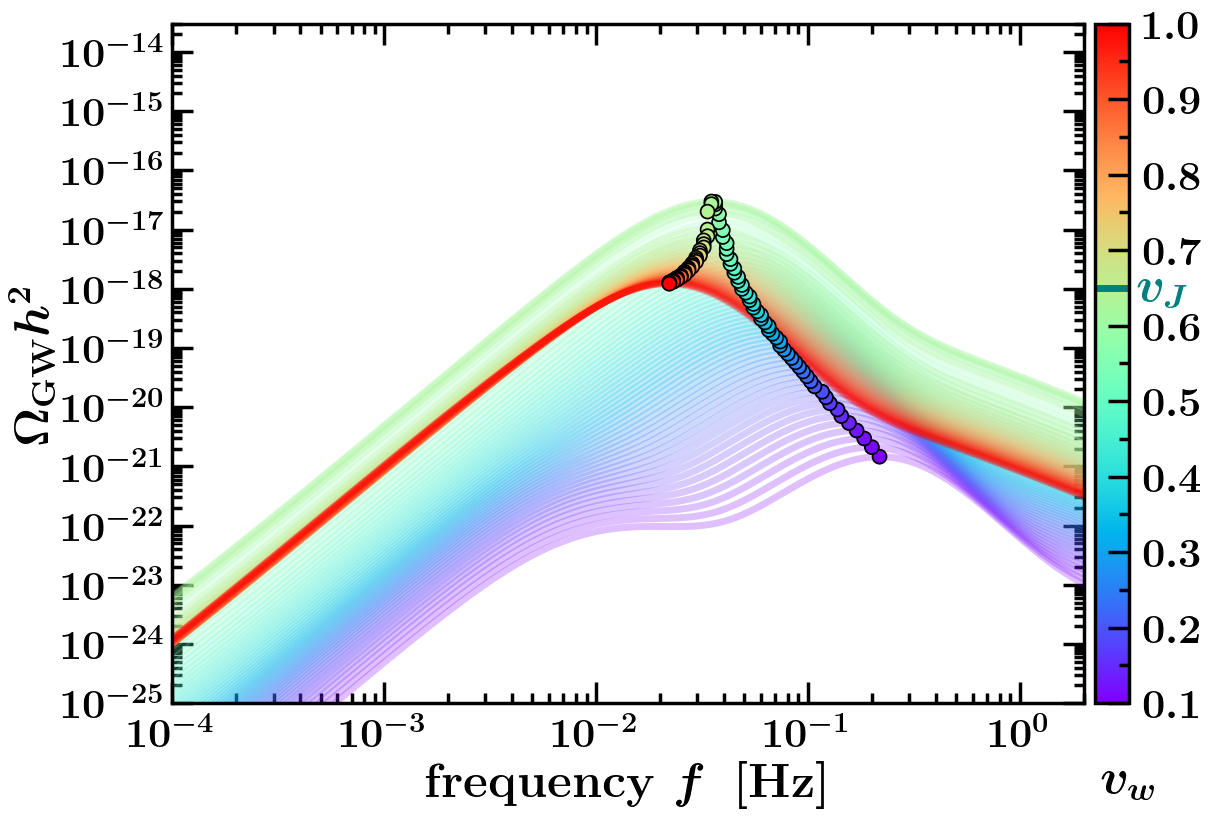}
        \caption{}
        \label{fig_vw_vary_bp1_sm}
    \end{subfigure}
    \begin{subfigure}{0.32\textwidth}
        \centering
        \includegraphics[width=1\linewidth]{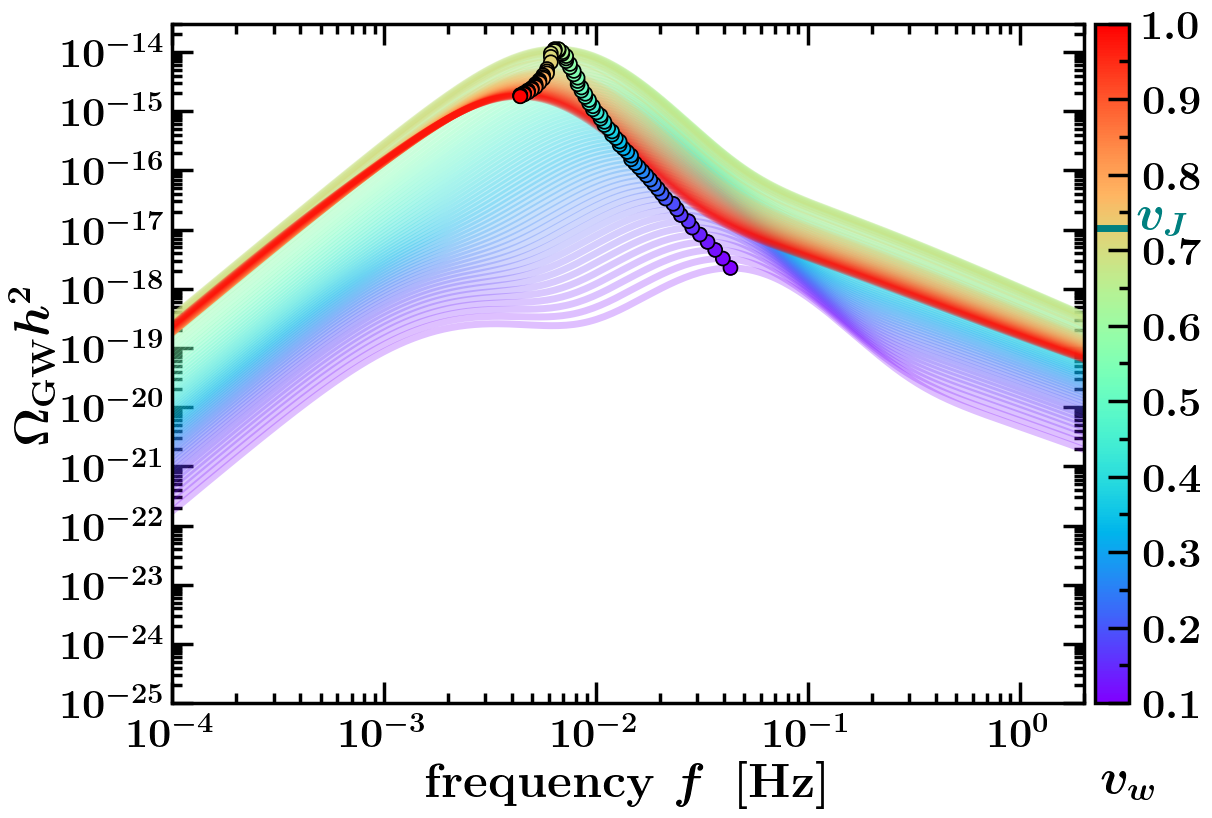}
        \caption{}
        \label{fig_vw_vary_bp1_mix}
    \end{subfigure}
	\caption{Variation of the GW spectrum with respect to the wall velocity for BP1, BP4 and BP5 as shown in the left (\ref{fig_vw_vary_bp1_bsm}), middle (\ref{fig_vw_vary_bp1_sm}) and right (\ref{fig_vw_vary_bp1_mix}) panels respectively. The locus of the peak frequency and the peak amplitude in the GW spectrum is encircled in black while the Chapman-Jouguet velocity is denoted by  teal horizontal line in the colorbar.}
	\label{fig_vw_vary}
\end{figure}

\noindent
In this appendix, we present the variation of the bubble wall velocity and its impact on the GW spectrum for three representative benchmark points as shown in Fig. \ref{fig_vw_vary}. Depending on the value of the $v_w$, the profile of the expanding fluid falls in three possible modes i.e., deflagration ($v_w \leq c_s$), hybrid ($c_s < v_w \leq v_J$) and detonation $(v_w > v_J)$ where $c_s = 1/\sqrt{3}$ is the speed of sound and $v_J$ is the Chapman-Jouguet velocity \cite{Steinhardt:1981ct, Espinosa:2010hh} given by
\begin{equation}
    v_J = \dfrac{1}{1 + \alpha} \left( \dfrac{1}{\sqrt{3}} + \sqrt{\alpha^2 + \dfrac{2 \alpha}{3}} \right) \ ,
\end{equation}
where $\alpha$ measures the transition strength in terms of the released vacuum energy. The peak frequency of the spectrum falls off with increasing wall velocity as can be inferred from Eqs. \ref{eq_peak_freq_coll}, \ref{eq_peak_freq_sw} and \ref{eq_peak_freq_turb}. We also note an opposite trend in the variation of the peak amplitude for the deflagration/hybrid and detonation regime with respect to the wall velocity consistent with detailed numerical simulations \cite{Tian:2024ysd}.

\end{appendix}


\bibliographystyle{JHEP}
\bibliography{cxSM_phase_transition_v2}
\end{document}